\documentstyle[preprint,aps]{revtex} 
 
\begin{document} 
 
\title{Stochastic boundary conditions in the 
deterministic Nagel-Schreckenberg traffic model} 
 
\author{S. Cheybani$^{a,b}$ and J.Kert\'{e}sz$^{b,c}$ and M. 
Schreckenberg$^{a}$} 
 
\address{$^a$ 
Theoretische Physik, Gerhard-Mercator Universit\"at, D-47048 Duisburg, Germany} 
\address{$^b$ 
Department of Theoretical Physics, Technical University of Budapest, H-1111 
Budapest, Hungary} 
\address{$^c$ 
Laboratory of Computational Engineering, Helsinki University of Technology, 
FIN-02150 Espoo, Finland} 
 
\maketitle 
\begin{abstract} 
\noindent 
We consider open systems where cars move according to the deterministic 
Nagel-Schreckenberg rules \cite{Na:Schr} and with maximum velocity 
${\mbox{v}}_{max}$ $>$ 1, what is an extension of the Asymmetric Exclusion 
Process (ASEP). 
It turns out that the behaviour of the system is dominated by two features: 
a) the competition between the left and the right boundary 
b) the development of so-called ''buffers'' due to the hindrance an 
injected car feels from the front car at the beginning 
of the system. 
As a consequence, there is a first-order phase transition between the 
free flow and the congested phase accompagnied by the collapse of the buffers 
and the phase diagram essentially differs from that for 
${\mbox{v}}_{max}$ = 1 (ASEP). 
\end{abstract} 
\narrowtext 
 
\section{Introduction} 
\noindent 
Driven diffusive processes have been widely studied as prototypes of 
non-equilibrium systems (\cite{Zia:Sha:Schm:Ast}--\cite{Kr:Fer}). They are 
modelled as a lattice gas and are characterized by a constant external force 
(e. g. electrical field) which sets up a steady current transporting 
information from the boundaries to the bulk of the system.\\ 
A well-known modification of the basic one-dimensional diffusive system is 
the asymmetric exclusion process (ASEP) which was first solved by  by Derrida 
et al \cite{De:Do:Mu} for open boundary conditions. The ASEP is defined as 
follows: Consider a one-dimensional lattice of L sites. Each site i 
(1 $\le$ i $\le$ L) is either occupied by a particle (${\tau}_{i}$ = 1) 
or empty (${\tau}_{i}$ = 0). A particle on site i has the probability p 
of hopping one site to the right if site i+1 is empty. At the left boundary 
of the system a particle is injected with probability $\alpha$ if i = 1 is 
empty. At the right boundary a particle on i = L is removed with probability 
$\beta$. The ASEP can be divided into four classes according to the order 
in which to perform hopping, injection and removal: 
\begin{eqnarray} 
&& \mbox{(a) random - sequential update} 
(\cite{De:Do:Mu} - \cite{Schue:Do}) \nonumber \\ 
&& \mbox{(b) ordered - sequential update} 
(\cite{Ra:Schr}, \cite{Ra:Scha:Schr}) \nonumber \\ 
&& \mbox{(c) sublattice - parallel update} 
(\cite{Ra:Schr} - \cite{Hon:Pesch}) \nonumber \\ 
&& \mbox{(d) parallel update} 
(\cite{Ti:Er} - \cite{Be:Cha:Ez})\nonumber 
\end{eqnarray} 
A detailed overview over all update types is given in \cite{Ra:Sa:Scha:Schr}.\\ 
An interesting feature of the ASEP is that phase transitions occur as a 
function of the model parameters. Usually there is a low density/high current 
phase and a high density/low current phase reminiscent to the ''free flow'' 
and the ''jamming'' states in vehicular traffic 
\cite{Wo:Schr:Ba}-\cite {Chowd:Sa:Scha}. Being a cellular automaton 
the ASEP and its generalizations are well-suited to serve as simple models for 
traffic problems since efficient analytical and numerical techniques have 
been developed for their study.\\ 
As it is common for traffic simulations we will use parallel update in the 
following because this is the most effective among the 
four update types and shows the best congruence with real traffic 
data \cite{Schr:Scha:Na:Ito}. In Fig 1a we reproduced the main results for 
the ASEP with parallel update: Based on investigations on global density, 
current, density profiles and correlation functions it turns out that there 
are two regimes, free flow and jamming, which are separated by the 
$\alpha$=$\beta$-line ($\alpha$: injection rate, $\beta$: extinction rate). 
All parameters have in common that they do not depend on the extinction 
rate $\beta$ (injection rate $\alpha$) in the free flow (jamming) regime.\\ 
Comparing the ASEP with real traffic, however, it is obvious that phenomena 
like acceleration and slowing down are not included in the model. Here, cars 
either do not move at all or move one site per time step. It can therefore be 
said, that they move with maximum velocity ${\mbox{v}}_{max}$ = 1. In order 
to get more realistic results, Nagel and Schreckenberg introduced a model 
\cite{Na:Schr}, where cars are able to drive with different discrete integer 
velocities v, 
0 $\le$ v $\le$ ${\mbox{v}}_{max}$ $>$ 1. 
\\ 
Another interesting feature of the parallel update procedure is that it 
induces additional short-range correlations 
compared to other updating procedures. An essential part of this paper 
will be therefore devoted to the investigation of short-range correlation 
functions (Section V) which have been already studied in corresponding systems 
with periodic boundary conditions for ${\mbox{v}}_{max}$ $\ge$ 1 
\cite{Sas:Kert} - \cite{Sa:Scha} and in systems with open boundary conditions 
for ${\mbox{v}}_{max}$ = 1 \cite{Ev:Ra:Spe}. 
Moreover, it turned out that correlation functions are well suited 
to describe the free flow - jamming transition \cite{Sas:Kert} - 
\cite{Ro:Lue:Us}. \\ 
The most significant difference between systems with open and periodic 
boundary conditions is the car density $\rho$. In a periodic system the 
car density is a tuning parameter and the probability to find a car at 
a certain site i is $\rho$. In systems with open boundary conditions, 
however, the situation is different as we have to deal with two 
different tuning parameters, namely the injection rate $\alpha$ and the 
extinction rate $\beta$ and the density is a derived parameter. 
\\ 
The influence of $\alpha$ and $\beta$ on the car density implies that 
quantities like global density, current, and the density profile, which 
were studied for the ASEP (${\mbox{v}}_{max}$ = 1) in \cite{Schr:Scha:Na:Ito}, 
\cite{Scha1:Schr1} - \cite{Scha}, show a different behaviour than in periodic 
systems.  For the case ${\mbox{v}}_{max}$ $>$ 1 and open boundary conditions, 
however, only few results exist. 
Therefore, the cases ${\mbox{v}}_{max}$ = 1 and 
${\mbox{v}}_{max}$ $>$ 1 in systems with open boundary 
conditions will be compared with each other in this paper, too.\\ 
The paper is organized as follows: 
In the next section the model is described. The current and the global 
density of the system are considered in Section III, in 
particular for the cases $\beta$ = 1, $\alpha$ = 1, and $\beta$ = 1-$\alpha$. 
In Section IV we analyze the corresponding density profiles and in Section V 
the short-range correlation functions. The results are summarized and 
discussed in Section VI.\\ 
\section{Model} 
\noindent 
Our investigations are based on a one-dimensional 
probabilistic cellular automaton model introduced by Nagel and 
Schreckenberg \cite{Na:Schr}. According to the Nagel - Schreckenberg (NS) 
model, the road is divided into L cells of equal size and the time is 
also discrete. Each site can be either empty or 
occupied by a car with velocity v = 0, 1, $\dots$, 
${\mbox{v}}_{max}$. All sites are simultaneously 
updated according to four successive steps: 
\begin{eqnarray} 
&& \mbox{1. Acceleration: increase v by 1 if v $<$ 
         ${\mbox{v}}_{max}$.} 
\nonumber \nonumber \\ 
&& \mbox{2. Slowing down: decrease v to v = d if necessary 
         (d: number of empty cells in front of the car).} 
\nonumber \nonumber \\ 
&& \mbox{3. Randomization: decrease v by 1 with randomization 
         probability p if p $>$ 0.} 
\nonumber \nonumber \\ 
&& \mbox{4. Movement: move car v sites forward.} 
\nonumber 
\end{eqnarray} 
It is obvious that the NS - model is identical with the ASEP model 
with parallel update for maximum velocity 
${\mbox{v}}_{max}$ = 1. In this paper, the randomization probability 
is p = 0, i. e. step 3 (randomization) is ignored. The investigations are 
mainly focused on ${\mbox{v}}_{max}$ = 5 but for comparison also 
maximum velocities ${\mbox{v}}_{max}$ = 2, 3, 4, 6, 7, ... are considered 
(see Section III).\\ 
Open boundary conditions are defined in the following way:\\ 
The system consists of L sites i with 1 $\le$ i $\le$ L (for the numerical 
simulations: L = 1024). At site i = 0, that means out of the system a vehicle 
with the 
probability $\alpha$ and with the velocity v = ${\mbox{v}}_{max}$ is created. 
This car immediately moves according to the NS rules. If i = 1 is occupied by 
another car so that the velocity of the injected car on i = 0 is v = 0 
then the injected car is deleted. At i = L+1 a ''block'' 
occurs with probability 1 - $\beta$ and causes a slowing down of the cars at 
the end of the system. Otherwise, with probability $\beta$, the cars simply 
move out of the system.\\ 

\section{Current and Global Density} 
\noindent 
The phase diagrams for systems with maximum velocities 
${\mbox{v}}_{max}$ = 2, 3, 5 are shown in Figs 1b-d. Fig 1b resembles the case 
${\mbox{v}}_{max}$ = 1 except for some deviations which are due to the fact 
that in systems with ${\mbox{v}}_{max}$ = 2 we do not have a particle-hole 
symmetry as for ${\mbox{v}}_{max}$ = 1. The course of the free flow - jamming 
border for the case ${\mbox{v}}_{max}$ = 3, on the other hand, is very 
different (Fig 1c). Here, the $\alpha$ = $\beta$ - line does not separate the 
free flow and the jamming regime. Instead, the jamming regime is larger than 
the free flow regime, and for high extinction rates $\beta$ cars freely move 
for $all$ $\alpha$. For the maximum velocity ${\mbox{v}}_{max}$ = 5 these 
features are even stronger developed as it is obvious from Fig 1d.\\ 
\\ 
Let us have a closer look at the $\beta$ = 1 - line. The current q in Fig 2a 
increases with increasing $\alpha$; for ${\mbox{v}}_{max}$ $\ge$ 5 we have 
q($\alpha$ $\le$ 0.5, $\beta$ = 1) = $\alpha$. For high injection rates, 
however, the curves surprisingly decrease (if ${\mbox{v}}_{max}$ $\ge$ 4). 
This phenomenon cannot be observed in systems with maximum velocities 
${\mbox{v}}_{max}$ = 2, 3 and for ${\mbox{v}}_{max}$ = 4 it is extremely weak. 
The maximum of the current is at $\alpha$ $\approx$ 0.9 for 
${\mbox{v}}_{max}$ = 5 and at $\alpha$ $\approx$ 0.835 for higher maximum 
velocities.\\ 
The corresponding global density $\overline{\rho}$ results from the current in Fig
 2a by the relation 
\begin{eqnarray} 
\overline{\rho} ( \alpha, \beta = 1) = 
\frac{q ( \alpha, \beta = 1)}{{\mbox{v}}_{max}} 
\nonumber 
\end{eqnarray} 
as all cars freely move with maximum velocity ${\mbox{v}}_{max}$.\\ 
\\ 
Considering the current (Fig 2b) and the global density (Fig 2c) for the 
injection rate $\alpha$ = 1, we see that for ${\mbox{v}}_{max}$ = 2 these 
quantities behave similarly to the case ${\mbox{v}}_{max}$ = 1. For 
${\mbox{v}}_{max}$ $\ge$ 3 astonishing effects are observed which do not 
depend on the maximum velocity if ${\mbox{v}}_{max}$ $\ge$ 5: Coming from 
low extinction rates $\beta$ the current for ${\mbox{v}}_{max}$ $\ge$ 5 
increases proportionally to $\beta$ and abruptly becomes constant at 
${\beta}_{c}$ = 0.835. For the global density, on the other hand, the 
transition seems to be continuous.\\ 
Investigations of systems for large system sizes, however, show that the 
continuous change in the global density is just a finite size effect: 
Although the curves are qualitatively the same as those in Fig 2c the 
transition from free flow to jamming becomes more and more abrupt with 
increasing system size L. Furthermore, it turns out that the value of 
${\beta}_{c}$ is slightly smaller than for L = 1024. As a consequence from 
numerical investigations of systems with large L it is fair to assume that for 
L $\rightarrow$ $\infty$ the current is described by 
\begin{eqnarray} 
\mbox{q}( \alpha = 1, \beta < \frac{5}{6}, {\mbox{v}}_{max} \ge 5) 
&& \,\ = \,\ 
\frac{4}{5} \beta \hspace{1cm} \mbox{jamming} \nonumber \\ 
\mbox{q}( \alpha = 1, \beta > \frac{5}{6}, {\mbox{v}}_{max} \ge 5) && \,\ 
= \,\ \frac{2}{3} \hspace{1.5cm} \mbox{free flow} \nonumber 
\end{eqnarray} 
and the corresponding global density is given by 
\begin{eqnarray} 
\overline{\rho} ( \alpha = 1, \beta < \frac{5}{6}, 
{\mbox{v}}_{max} \ge 5) &&  \,\ = \,\ 
1 - \frac{4}{5} \beta \hspace{1cm} \mbox{jamming} \nonumber \\ 
\overline{\rho} ( \alpha = 1, \beta > \frac{5}{6}, 
{\mbox{v}}_{max} \ge 5) && \,\ = \,\ 
\frac{2}{3 {\mbox{v}}_{max}} \hspace{1.5cm} \mbox{free flow} \nonumber 
\end{eqnarray} 
For increasing system sizes current and global density converge to these 
values which can be ``calculated'' analytically as it is demonstrated in the 
following. 
Unfortunately, there exist no extensive analytical theory of the NS model for 
maximum velocities ${\mbox{v}}_{max}$ $>$ 1. We must therefore restrict 
ourselves to a kind of bookkeeping which is nevertheless well-suited for 
the understanding of what is going on in the system. 
Furthermore, it should be emphasized that the representations of the 
configurations are snapshots between the slowing down step and the movement 
step. This is just a convention and does not change anything in the physical 
meaning. \\ 
\\ 
 In order to get a better insight in the behaviour of the current and the
 global density we consider the special case $\alpha$ = $\beta$ = 1. The car
 velocity is represented by numbers in brackets,
 (v) = (0), (1), ..., (${\mbox{v}}_{max}$), and k connected unoccupied sites
 by the symbol ${x}^{\mbox{k}}$. The first number in brackets represents
 the car at i = 0 where cars are injected. Then we have for
 \begin{eqnarray}
 && \mbox{t}=0: \,\ \,\  ({\mbox{v}}_{max}) \,\ {x}^{L} \nonumber \\
 && \mbox{t}=1: \,\ \,\   ({\mbox{v}}_{max}-1) \,\ {x}^{{\mbox{v}}_{max}-1} \,\
                        ({\mbox{v}}_{max}) {x}^{L-{\mbox{v}}_{max}}
 \nonumber \\
 && \mbox{t}=2: \,\ \,\  ({\mbox{v}}_{max}-2) \,\ {x}^{{\mbox{v}}_{max}-2} \,\
            ({\mbox{v}}_{max}) \,\
             {x}^{{\mbox{v}}_{max}} ({\mbox{v}}_{max}) \,\
            {x}^{L-2{\mbox{v}}_{max}} \nonumber \\
 &&  \hspace{4cm}   \vdots \nonumber
 \end{eqnarray}
 Now, a pair of consecutive cars is focussed found at the beginning of the
 system at time t = $n$ with 2 $\ge$ $n$ $\ge$ ${\mbox{v}}_{max}$ - 1:
 \begin{eqnarray}
 \mbox{t}=n: \,\ \,\  ({\mbox{v}}_{max}-n) \,\
                {x}^{{\mbox{v}}_{max}-\mbox{$n$}} \,\
            ({\mbox{v}}_{max}-n+2) \,\ \cdots
             ({\mbox{v}}_{max}) \,\
            {x}^{L-\mbox{$n$}{\mbox{v}}_{max}}. \nonumber
 \end{eqnarray}
 The difference of the velocities of the cars is
 $\Delta$v = ${\mbox{v}}_{front}$ - ${\mbox{v}}_{back}$ = 2 and the velocity
 of each car increases by 1 due to the acceleration step of the NS model.
 Consequently, the space between the cars grows with $\Delta$v t = 2t.
 After $n-1$ time steps we have
 \begin{eqnarray}
 \mbox{t}=2n-1: \,\ \,\  \cdots \,\ \,\
              ({\mbox{v}}_{max}-1) \,\
              {x}^{{\mbox{v}}_{max}+\mbox{$n$}-2} \,\
            ({\mbox{v}}_{max}) \,\ \cdots
             ({\mbox{v}}_{max}) \,\
            {x}^{L-(2\mbox{$n$}-1){\mbox{v}}_{max}} \nonumber
 \end{eqnarray}
 and finally, we find
 \begin{eqnarray}
 \mbox{t}=2n: \,\ \,\ \,\ \cdots \cdots \,\ \,\
              ({\mbox{v}}_{max}) \,\
               {x}^{{\mbox{v}}_{max}+\mbox{$n$}-1} \,\
            ({\mbox{v}}_{max}) \,\ \cdots
             ({\mbox{v}}_{max}) \,\
            {x}^{L-2\mbox{$n$}{\mbox{v}}_{max}}. \nonumber
 \end{eqnarray}
 From now on, the space between the front and the back cars keep constant and
 consists of maximally $2({\mbox{v}}_{max}-1)$ empty sites due to
 $n \le {\mbox{v}}_{max}-1$.\\
 \\
 The situation is different for the case $n = {\mbox{v}}_{max}$:
 \begin{eqnarray}
 \mbox{t}={\mbox{v}}_{max}: \,\ \,\
 (0)(2) \cdots ({\mbox{v}}_{max}) {x}^{L-{\mbox{v}}_{max}^2}
 \nonumber
 \end{eqnarray}
 According to the left boundary conditions the car at site i = 0 with velocity
 v = 0 is deleted and a new car is created instead at the next time step:
 \begin{eqnarray}
 \mbox{t}={\mbox{v}}_{max}+1: \,\ \,\
 (2) {x}^{2} (3) \cdots ({\mbox{v}}_{max})
 {x}^{L-{\mbox{v}}_{max}({\mbox{v}}_{max}+1)}
 \nonumber
 \end{eqnarray}
 Here, $\Delta$v = 1 and the space between the cars grows as $\Delta$v t = t.
 After ${\mbox{v}}_{max}$-2 time steps we finally get
 \begin{eqnarray}
 \mbox{t}=2{\mbox{v}}_{max}-1: \,\ \,\
 \cdots \cdots ({\mbox{v}}_{max}) {x}^{{\mbox{v}}_{max}} ({\mbox{v}}_{max})
 \cdots ({\mbox{v}}_{max})
 {x}^{L-{\mbox{v}}_{max}(2{\mbox{v}}_{max}-1)}
 \nonumber
 \end{eqnarray}
 If one proceeds, it can be clearly seen that there are three scenarios
 ($m = 0, 1, 2, \cdots$):
 The car created at site i = 0 and t = $n$
 (a) is deleted according to the left boundary conditions if
     $n = {\mbox{v}}_{max} + 3m$.
 (b) has ${\mbox{v}}_{max}$ empty sites in front if
     $n = {\mbox{v}}_{max} + 1 + 3m$
 (c) has $2({\mbox{v}}_{max}-1)$ empty sites in front if
     $n = {\mbox{v}}_{max} + 2 + 3m$.
 In other words, a self-repeating pattern establishes itself after a while
 according to
 \begin{eqnarray}
 && (2) \,\ {x}^{2} \,\ \,\ \cdots \,\ \,\ ({\mbox{v}}_{max}) \,\ {x}^{2({\mbox{v}}_{max}-1)} \,\
 ({\mbox{v}}_{max}) \,\ {x}^{{\mbox{v}}_{max}} \,\
 ({\mbox{v}}_{max}) \,\ {x}^{2({\mbox{v}}_{max}-1)} \,\
 ({\mbox{v}}_{max}) \,\ {x}^{{\mbox{v}}_{max}} \,\
 ({\mbox{v}}_{max}) \,\ {x}^{2({\mbox{v}}_{max}-1)} \,\
 \cdots \nonumber \\
 && (1) \,\ {x}^{1} \,\ \,\ \cdots \,\ \,\ ({\mbox{v}}_{max}) \,\ {x}^{2({\mbox{v}}_{max}-1)} \,\
 ({\mbox{v}}_{max}) \,\ {x}^{{\mbox{v}}_{max}} \,\
 ({\mbox{v}}_{max}) \,\ {x}^{2({\mbox{v}}_{max}-1)} \,\
 ({\mbox{v}}_{max}) \,\ {x}^{{\mbox{v}}_{max}} \,\
 ({\mbox{v}}_{max}) \,\ {x}^{2({\mbox{v}}_{max}-1)} \,\
 \cdots \nonumber \\
 && (0) \,\ (2) \,\ {x}^{2} \cdots \,\ \,\ ({\mbox{v}}_{max}) \,\ {x}^{2({\mbox{v}}_{max}-1)} \,\
 ({\mbox{v}}_{max}) \,\ {x}^{{\mbox{v}}_{max}} \,\
 ({\mbox{v}}_{max}) \,\ {x}^{2({\mbox{v}}_{max}-1)} \,\
 ({\mbox{v}}_{max}) \,\ {x}^{{\mbox{v}}_{max}} \,\
 ({\mbox{v}}_{max}) \,\ {x}^{2({\mbox{v}}_{max}-1)} \,\
 \cdots \nonumber
 \end{eqnarray}
 This is perhaps astonishing because we naively would expect ${\mbox{v}}_{max}$
 unoccupied sites between two neighbouring cars for $\alpha$ = 1. Actually,
 there are also spaces consisting of 2(${\mbox{v}}_{max}$-1) sites what is a
 consequence of the hindrance the injected cars feels from the front car at
 the beginning of the system. In other words, ${\mbox{v}}_{max}$-2 additional
 sites - so-called ''buffers'' (the motivation for this name
 will be explained later) - occur playing an important role for systems
 with maximum velocity ${\mbox{v}}_{max}$ $\ge$ 3 as we will see below.\\
 Besides, our reflections clearly show that one has to wait for at least
 t = ${\mbox{v}}_{max}$ time steps until the self-repeating pattern is
 established. Within this time period the first car created at t = 0 has moved
 onto site i = ${\mbox{v}}_{max}^2$. Therefore, our considerations are only
 valid for systems which size is much larger than ${\mbox{v}}_{max}^2$,
 otherwise, boundary effects must be taken into account.
 \\
 \\
 From the self-repeating pattern follows that the distance between two
 neighbouring cars driving with ${\mbox{v}}_{max}$ is alternately
 ${\mbox{d}}_{1}$ = ${\mbox{v}}_{max}$ and
 ${\mbox{d}}_{2}$ = 2(${\mbox{v}}_{max}$ - 1), i.e.,
 \begin{eqnarray}
 \mbox{${\mbox{d}}_{1}$ = 2, ${\mbox{d}}_{2}$ = 2 } \,\ \,\
 \mbox{for ${\mbox{v}}_{max}$ = 2} \nonumber \\
 \mbox{${\mbox{d}}_{1}$ = 3, ${\mbox{d}}_{2}$ = 4 } \,\ \,\
 \mbox{for ${\mbox{v}}_{max}$ = 3} \nonumber \\
 \mbox{${\mbox{d}}_{1}$ = 4, ${\mbox{d}}_{2}$ = 6 } \,\ \,\
 \mbox{for ${\mbox{v}}_{max}$ = 4} \nonumber \\
 \mbox{${\mbox{d}}_{1}$ = 5, ${\mbox{d}}_{2}$ = 8 } \,\ \,\
 \mbox{for ${\mbox{v}}_{max}$ = 5} \nonumber \\
 \mbox{${\mbox{d}}_{1}$ = 6, ${\mbox{d}}_{2}$ = 10 } \,\
 \mbox{for ${\mbox{v}}_{max}$ = 6} \nonumber \\
 \vdots \hspace*{5.1cm} \vdots \nonumber
 \end{eqnarray}
 That means, buffers occur only for ${\mbox{v}}_{max}$ $\ge$ 3.\\
 ${\mbox{v}}_{max}$ = 2 is a special case behaving similarly to
 ${\mbox{v}}_{max}$ = 1. It is therefore no surprise that the corresponding
 phase diagram, the global density, and the current resembles the case
 ${\mbox{v}}_{max}$ = 1.
 If finite size effects are left out of consideration the current is
 obviously given by
 \begin{eqnarray}
 \mbox{q}(\alpha = \beta = 1, {\mbox{v}}_{max} > 1) = \frac{2}{3} \nonumber
 \end{eqnarray}
 and the global density by
 \begin{eqnarray}
 \overline{\rho}(\alpha = \beta = 1, {\mbox{v}}_{max} > 1) =
 \frac{2}{3 {\mbox{v}}_{max}} \nonumber
 \end{eqnarray}
 what coincides with numerical results.\\
 \\
 We will now investigate the effect of the buffers for the extinction rate
 $\beta$ = 1. For that purpose we consider a slightly smaller injection rate
 by working a ''disturbance'' in the $\alpha$ = $\beta$ = 1 pattern, i. e.,
 at each time step {\it except for one} a car is created at i = 0. As the
 self-repeating pattern consists of three time steps we have three
 possibilities  to place the disturbance. In Appendix A the effect is
 illustrated for systems with ${\mbox{v}}_{max}$ $\ge$ 5 (because of lack of
 space we set \^v $\equiv$ ${\mbox{v}}_{max}$ in Appendices A, B). It turns
 out that the movement of the cars does not change at all for possibility (c).
 For (a) and (b), however, the disturbance influences the system for three
 time steps as three cars show a deviating behaviour in Appendix A. Having a
 closer look on the sites affected by the disturbance we see that the current
 ${\mbox{q}}_{dist}$($\beta$ = 1, ${\mbox{v}}_{max}$ $\ge$ 5) = $\frac{3}{4}$
 and the global density
 ${\overline{\rho}}_{dist}$($\beta$ = 1, ${\mbox{v}}_{max}$ $\ge$ 5) =
  $\frac{3}{4 {\mbox{v}}_{max}}$
 are higher there than for $\alpha$ = $\beta$ = 1. As altogether
 4${\mbox{v}}_{max}$ sites are concerned by the disturbance the effect
 increases with increasing ${\mbox{v}}_{max}$.\\
 Considering the site i = 0 in Appendix A it is obvious that the effect of
 the disturbance is different for maximum velocities ${\mbox{v}}_{max}$ $<$ 5
 as cars driving with ${\mbox{v}}_{max}$ = 4 cannot be injected with v = 5,
 cars driving with ${\mbox{v}}_{max}$ = 3 not with v = 4 and so on. We do not
 go into details but just list up the results: Placing a disturbance at the
 beginning of a system with ${\mbox{v}}_{max}$ = 2, 3, 4 one gets
 \begin{eqnarray}
 \mbox{(a)} \,\ \,\ &&
 {\mbox{q}}_{dist} ( \beta = 1, {\mbox{v}}_{max} = 4) =  \frac{12}{17},
 \,\ \,\
 {\overline{\rho}}_{dist} ( \beta = 1, {\mbox{v}}_{max} = 4) =
 \frac{3}{17}
 \nonumber \\
 &&
 {\mbox{q}}_{dist} ( \beta = 1, {\mbox{v}}_{max} = 3) =  \frac{1}{2},
 \,\ \,\ \,\
 {\overline{\rho}}_{dist} ( \beta = 1, {\mbox{v}}_{max} = 3) =
  \frac{1}{6}
 \nonumber \\
  &&
 {\mbox{q}}_{dist} ( \beta = 1, {\mbox{v}}_{max} = 2) =  \frac{2}{5},
 \,\ \,\ \,\
 {\overline{\rho}}_{dist} ( \beta = 1, {\mbox{v}}_{max} = 2) =
  \frac{1}{5}
 \nonumber \\
  \mbox{(b)} \,\ \,\ &&
 {\mbox{q}}_{dist} ( \beta = 1, {\mbox{v}}_{max} = 4) =  \frac{3}{4},
 \,\ \,\ \,\
 {\overline{\rho}}_{dist} ( \beta = 1, {\mbox{v}}_{max} = 4) =
  \frac{3}{16}
 \nonumber \\
  &&
 \mbox{no effect of disturbance for ${\mbox{v}}_{max}$ = 3}
 \nonumber \\
  &&
 {\mbox{q}}_{dist} ( \beta = 1, {\mbox{v}}_{max} = 2) =  \frac{1}{2},
 \,\ \,\ \,\
 {\overline{\rho}}_{dist} ( \beta = 1, {\mbox{v}}_{max} = 2) =
  \frac{1}{4}
 \nonumber \\
  \mbox{(c)} \,\ \,\ &&
  \mbox{no effect of disturbance for ${\mbox{v}}_{max}$ = 2, 3, 4}
 \nonumber
 \end{eqnarray}
 Superposition of all possibilities (a), (b), and (c) leads to the result that
 the effect of the disturbance is weaker for ${\mbox{v}}_{max}$ = 4 than for
 corresponding systems with ${\mbox{v}}_{max}$ $\ge$ 5. For maximum velocities
 ${\mbox{v}}_{max}$ = 2, 3 the current and the global density decrease and
 that is why the maximum of the curves in Fig 2a is at $\alpha$ = 1 if
 ${\mbox{v}}_{max}$ $\le$ 3.\\
 As far as the position of the maximum of the current is concerned we can
 only give a hand-waving argument: It is obvious from Appendix A that the
 disturbance affects the development of two buffers. On the other hand, it
 can be easily seen that for $\alpha$ = $\beta$ = 1 a buffer is created every
 three time steps (and consequently, two buffers are created in six time steps)
. Therefore, the strongest effect is expected when the system is disturbed
 with the rate (1-$\alpha$) = $\frac{1}{6}$. If (1-$\alpha$) becomes higher
 the buffers being necessary for the increase in the current and the global
 density cannot develop. This may be the reason why the maximum for the curves
 in Fig 2a with ${\mbox{v}}_{max}$ $>$ 5 is found at $\alpha$ $\approx$
 $\frac{5}{6}$.\\
 \\
 For the injection rate $\alpha$ = 1 the buffers have an even more dramatic
 effect which can be observed at the end of the system.In analogy to
 $\beta$ = 1 we start with the special case $\alpha$ = $\beta$ = 1.
 By simple analytic considerations it turns out that a self-repeating
 pattern
 \begin{eqnarray}
 && \cdots \,\ \,\  {x}^{2({\mbox{v}}_{max}-1)} \,\
 ({\mbox{v}}_{max}) \,\ {x}^{{\mbox{v}}_{max}} \,\
 ({\mbox{v}}_{max}) \,\ {x}^{2({\mbox{v}}_{max}-1)} \,\
 ({\mbox{v}}_{max}) \,\ {x}^{{\mbox{v}}_{max}} \,\
 ({\mbox{v}}_{max}) \,\ {x}^{2({\mbox{v}}_{max}-1)} \,\
 ({\mbox{v}}_{max})
 \nonumber \\
 && \,\ \,\ \cdots \,\ \,\
 ({\mbox{v}}_{max}) \,\ {x}^{2({\mbox{v}}_{max}-1)} \,\
 ({\mbox{v}}_{max}) \,\ {x}^{{\mbox{v}}_{max}} \,\
 ({\mbox{v}}_{max}) \,\ {x}^{2({\mbox{v}}_{max}-1)} \,\
 ({\mbox{v}}_{max}) \,\ {x}^{{\mbox{v}}_{max}} \,\
 ({\mbox{v}}_{max}) \,\ {x}^{{\mbox{v}}_{max}-1}
 \nonumber \\
 && \,\ \,\ \,\ \cdots \,\ \,\
 ({\mbox{v}}_{max}) \,\ {x}^{{\mbox{v}}_{max}} \,\
 ({\mbox{v}}_{max}) \,\ {x}^{2({\mbox{v}}_{max}-1)} \,\
 ({\mbox{v}}_{max}) \,\ {x}^{{\mbox{v}}_{max}} \,\
 ({\mbox{v}}_{max}) \,\ {x}^{2({\mbox{v}}_{max}-1)}
 ({\mbox{v}}_{max}) \,\ {x}^{{\mbox{v}}_{max}}
 \nonumber
 \end{eqnarray}
 establishes itself at the end of the system, too
 (L $\gg$ ${\mbox{v}}_{max}^2$).
 It is important to mention that - due to $\beta$ = 1 - no blockage occurs at
 all at the right boundary and that the buffers reach the right boundary with
 the rate ${\alpha}_{buffer}$ = $\frac{1}{3}$. The introduction of a
 disturbance (i.e. the consideration of an extinction rate which is slightly
 smaller than $\beta$ = 1) means here to place a single blockage at the end of
 the system. According to the self-repeating pattern consisting of three time
 steps we have to consider three possibilities. From Appendix B it turns out
 that the ${\mbox{v}}_{max}$-2 additional sites (resulting from the hindrance
 the cars feel at the beginning of the system from the front car) play an
 important role at the end of the system, too. Here, they have the effect of a
 ''buffer'' against the influence of the right boundary.
 It can be seen from Appendix B that two buffers are necessary to neutralize
 the blockage effect at the end of the system. Therefore, as long as
 (1-$\beta$) $<$ $\frac{1}{2} {\alpha}_{buffer}$ = $\frac{1}{6}$ a jamming wave
 cannot develop.\\
 For ${\beta}_{c}$ = $\frac{5}{6}$, however, there is a jump in the global
 density (remember that our analytical considerations are based on systems with
 size L $\rightarrow$ $\infty$, for L = 1024 the change from free flow to
 traffic is less abrupt due to finite size effects). As mentioned above we have
 $\overline{\rho}$($\alpha$ = 1, $\beta$ $>$ ${\beta}_{c}$,
 ${\mbox{v}}_{max}$ $\ge$ 5) = $\frac{2}{3 {\mbox{v}}_{max}}$ in the free
 flow regime and $\overline{\rho}$($\alpha$ = 1, $\beta$ $<$ ${\beta}_{c}$,
 ${\mbox{v}}_{max}$ $\ge$ 5) = 1 - 0.8$\beta$ in the jamming regime.
 At ${\beta}_{c}$ = $\frac{5}{6}$ $\overline{\rho}$ immediately increases from
 $\frac{2}{3 {\mbox{v}}_{max}}$ (free flow) to $\frac{1}{3}$ (jamming). That
 means that there is a jump of $\frac{{\mbox{v}}_{max}-2}{3 {\mbox{v}}_{max}}$
 at the critical extinction rate what corresponds to the buffer density in the
 free flow regime. In other words: At ${\beta}_{c}$ = $\frac{5}{6}$ the
 buffers cannot neutralize the blockage at the right boundary any longer.
 The buffer effect breaks down, jamming waves propagate from the end of the
 system to the left, and the buffers (${\mbox{v}}_{max}$ - 2 sites on
 3${\mbox{v}}_{max}$ sites each) are completely occupied by cars. Consequently,
 both current and global density show similar behaviour as the corresponding
 quantities for ${\mbox{v}}_{max}$ = 1 if $\beta$ $<$ $\frac{5}{6}$.\\
 \\
 Another interesting feature observed in Figs 2b,c is that current and global
 density do not depend on the right (left) boundary conditions, i.e. not on
 $\beta$ (not on $\alpha$ and ${\mbox{v}}_{max}$), if the system is in the
 free flow (jamming) regime. This is not only valid for $\alpha$ = 1 but
 also for general injection and extinction rates as it can be seen in Fig 3a
 for the current and for Fig 3b for the global density.\\
 \\
 \\
 In order to compare our results with those for corresponding systems with
 periodic boundary conditions we investigate the case $\beta$ = 1-$\alpha$.
 For $\beta$ = 1-$\alpha$ there are rather similar conditions at the left
 and at the right boundary and therefore, systems with open and with periodic
 boundary conditions can be compared at best with each other.\\
 The fundamental diagram for systems with periodic boundary conditions (PBC)
 is completely determined by the maximum velocity ${\mbox{v}}_{max}$ (see
 \cite{Scha} and references therein). The current of the system is given by
 ${\mbox{q}}^{PBC}$($\rho$ $<$ ${\rho}_{c}$) = ${\mbox{v}}_{max} \rho$ for
 freely moving and by
 ${\mbox{q}}^{PBC}$($\rho$ $>$ ${\rho}_{c}$) = 1 - $\rho$ for jammed cars.
 The critical density is given by
 ${\rho}_{c}$ = $\frac{1}{{\mbox{v}}_{max} + 1}$.\\
 In the case of open boundary conditions, on the other hand, it turns out
 from numerical results for ${\mbox{v}}_{max}$ $\ge$ 5 that the current in the
 free flow (jamming) regime only depends on the injection (extinction) rate
 according to
 \begin{eqnarray}
 \mbox{q($\beta$ = 1-$\alpha$) = $\alpha$} \hspace{0.9cm}
 \mbox{for $\alpha$ $\le$ ${\alpha}_{c}$, $\beta$ $\ge$ ${\beta}_{c}$}
 \nonumber \\
 \mbox{q($\beta$ = 1-$\alpha$) = 0.8$\beta$} \,\ \,\
 \mbox{for $\alpha$ $\ge$ ${\alpha}_{c}$, $\beta$ $\le$ ${\beta}_{c}$}
 \nonumber
 \end{eqnarray}
 and consequently, the transition takes place at ${\alpha}_{c}$ = 0.44
 (${\beta}_{c}$ = 0.56).
 The global density for $\beta$ = 1-$\alpha$ shows finite size effects
 as in the case $\alpha$ = 1. For large L, however, the transition from free
 flow to jamming becomes sharp. Then the global density is described by
 \begin{eqnarray}
 \mbox{$\overline{\rho}$($\beta$ = 1-$\alpha$) =
 $\frac{\alpha}{{\mbox{v}}_{max}}$}
 \hspace{1cm}
 \mbox{for $\alpha$ $<$ ${\alpha}_{c}$, $\beta$ $>$ ${\beta}_{c}$}
 \nonumber \\
 \mbox{$\overline{\rho}$($\beta$ = 1-$\alpha$) = 1 - 0.8$\beta$} \,\ \,\
 \mbox{for $\alpha$ $>$ ${\alpha}_{c}$, $\beta$ $<$ ${\beta}_{c}$}
 \nonumber
 \end{eqnarray}
 with a jump at ${\alpha}_{c}$ = $\frac{4}{9}$ $\approx$ 0.44
 (${\beta}_{c}$ = $\frac{5}{9}$ $\approx$ 0.56).\\
 The results for $\beta$ = 1-$\alpha$ induce the identity
 \begin{eqnarray}
 \mbox{q($\beta$ = 1-$\alpha$) = q($\beta$ = 1)} \,\ \,\
 \mbox{for $\alpha$ $<$ ${\alpha}_{c}$, $\beta$ $>$ ${\beta}_{c}$} \nonumber \\
 \mbox{q($\beta$ = 1-$\alpha$) = q($\alpha$ = 1)} \,\ \,\
 \mbox{for $\alpha$ $>$ ${\alpha}_{c}$, $\beta$ $<$ ${\beta}_{c}$} \nonumber
 \end{eqnarray}
 This indicates that the movement of the
 vehicles in the high density or jamming regime is dominated by the right
 boundary conditions, in the low density or free flow regime by the left
 boundary conditions. To get a better insight in this question we will have a
 closer look on the density profiles and the short-range
 correlation functions which are analyzed due to the three special cases
 (see also Fig 1d)
 \begin{eqnarray}
 && \mbox{1. $\beta$ = 1: shows the influence of the left boundary}
 \nonumber \nonumber \\
 && \mbox{2. $\alpha$ = 1: shows the influence of the right boundary}
 \nonumber \nonumber \\
 && \mbox{3. $\beta$ = 1-$\alpha$: systems with open and periodic
 boundary conditions can be compared at best}
 \nonumber \nonumber\\
 && \mbox{\,\ \,\ \,\  with each other}
 \nonumber
 \end{eqnarray}
 \\
 The investigations of this section clearly show that the case
 ${\mbox{v}}_{max}$ = 5 includes all features which are characteristic for
 higher maximum velocities, too. For this reason we confine ourselves to
 systems with ${\mbox{v}}_{max}$ = 5 (and L = 1024) in the following.\\
 \\

 \section{Density Profiles}
 \subsection{$\beta$ = 1}
 \noindent
 In this section we investigate the influence of the left boundary on the
 density profiles. The best way to do this is to set $\beta$ = 1, because
 in that case the right boundary has no influence on the system.\\
 From Figs 4a,b it can be seen that the density profiles are characterized by
 a periodical structure. This is a significant difference to the case
 ${\mbox{v}}_{max}$ = 1 where oscillations cannot be found
 at all \cite{Ra:Sa:Scha:Schr}. For ${\mbox{v}}_{max}$ = 5,
 however, the density profiles show a certain pattern recurring with the
 period $\Delta$i = 5. In order to understand this phenomenon we consider
 the density profiles for very low injection rates first.\\
 For $\alpha$ = 0.05 (see Figs 4a,b) the probability of generating a car at
 i = 0 in two successive time steps is very small and therefore, the cars at
 the beginning of the system do not interact with each other. That means
 that a car which is created on i = 0 with velocity 5 (according to the left
 boundary conditions) moves to i = 5 at the next time step and can be found
 on the site i = 5n after n time steps (n = 1, 2, 3, ...). The density on these
 sites is $\rho$ $\approx$ $\alpha$. As it is obvious from Figs 4a,b a car
 can be also found on i = 5n+4 for small $\alpha$, too, but the probability
 for that is very small.\\
 For increasing injection rates $\alpha$, however, the probability of
 generating cars in two successive time steps increases and with it the
 hindrance a car at the beginning of the system feels from the front car.
 This can be understood as follows: Let us create a car A at time step t and
 a car B at time step t+1. Considering the system at t+1 we see that car A
 is on i = 5 having the velocity 5 whereas car B on i = 0 has the velocity 4
 because there are only four empty sites to car A. At the time step t+n, car
 A is on i = 5n and car B on i = 5(n-1)-1.\\
 To sum it up it can be said that the hindrance due to the left boundary
 conditions leads to a shift of the position of the cars within the system.
 This shift is reflected in the periodic pattern of Figs 4a,b. Whereas it is
 rather probable to find a car on i = 5n+5 and on i = 5n+4, the probability of
 finding a car on i = 5n+2 is much smaller and for i = 5n+3 it nearly
 vanishes.\\
 The most important result, however, is the fact that the sites i = 6+5n
 are {\it never} occupied according to the left boundary conditions so that
 the density on these sites have the value $\rho$(i=6+5n) = 0 for all $\alpha$.
 Before turning back to this point we have a  look at the case
 $\alpha$ = $\beta$ = 1 which is of special interest in the following section,
 too.\\
 For $\alpha$ = $\beta$ = 1 the corresponding density profile has the
 following form:
 \begin{eqnarray}
 && \mbox{$\rho$(i) = $\frac{1}{3}$ \,\ \,\ if i = 5n+4 or i = 5n+5}
 \nonumber \\
 && \mbox{$\rho$(i) = 0  \,\ \,\ else} \nonumber
 \end{eqnarray}
 as it can be easily deduced from the left boundary conditions.\\
 \subsection{$\alpha$ = 1}
 \noindent
 We investigate the influence of the right boundary now. Unfortunately, the
 influence of the left boundary cannot be completely left out of consideration
 by setting, for example, $\alpha$ = 0, because in that case no cars would be
 generated at all. Instead, we choose $\alpha$ = 1, because only for
 $\alpha$ = 1, the cars are deterministically created. This allows us to
 distinguish between the influence of the right and of the left boundary.\\
 In Fig 5a we can see that the situation for $\alpha$ = 1 is very different
 from that described in the previous section. For high extinction rates we
 still recognize the periodic structure already known from the case
 $\beta$ = 1. For extinction rates $\beta$ between 0.75 and 0.85 something
 interesting happens: the oscillations vanish and the envelope of the density
 profile rises. For low extinction rates the density profiles are just a
 constant which value increases with decreasing $\beta$.\\
 In order to understand this change we consider density profiles for
 0.83 $\le$ $\beta$ $\le$ 0.84 in Figs 5b-d. On i = 4+5n and i = 5+5n we find
 $\rho$(i) = $\frac{1}{3}$ resulting from the influence of the left boundary
 (see Section IV.A). The other sites (with $\rho$ = 0 for $\beta$ = 1),
 however, increasingly reflect the influence of the right boundary with
 decreasing extinction rates. Coming from high $\beta$ the density on i = 6+5n,
 i = 7+5n and i = 8+5n seems to ''come away'' from the $\rho$(i) = 0 - line
 starting at the right boundary.
 This phenomenon can be explained due to the repulsion the car feels at the
 right boundary with decreasing probability $\beta$ of being extincted.
 Consequently, a jam develops at the right boundary which expands to the left.
 For $\beta$ $\approx$ 0.837 the influence of the right boundary finally
 reaches the beginning of the system (Fig 5c). For $\beta$ = 0.84 the sites
 i = 4+5n and i = 5+5n indicate the repulsion at the right boundary, too, as
 the density profile becomes $\rho$ $>$ $\frac{1}{3}$ there. In parallel to
 this the oscillations resulting from the left boundary conditions vanish, a
 process which starts from the end of the system as well.\\
 \\
 Our observations have been quite qualitative so far. In the following the
 transition described above will be analyzed in detail and for that purpose
 we will have a closer look at the sites i = 6+5n. As we know from the
 previous section these sites are never occupied according to the left
 boundary conditions. In other words: The occupation of the sites i = 6+5n
 is {\it exclusively} an effect of the right boundary. Therefore, these sites
 play an important role as they show the repercussion of the right boundary
 on the system.\\
 The density on these sites is shown in Figs 6a-c. The density profiles
 correspond to the same $\beta$ as in Fig 5b but here, all sites except for
 i = 6+5n are left out of consideration.
 Let us first consider the density profiles for $\beta$ $>$ ${\beta}_{c}$
 (Fig 6b) which are exponential functions $\rho$ (i = 6+5n) =
 ${\rho}_{max}$( $\beta$ ) ${\mbox{e}}^{ \mbox{c}( \beta ) \mbox{(i-L)}}$
 (${\rho}_{max}$( $\beta$ ): maximum value of the density on the sites
 i = 6+5n). In Fig 6d the exponent c($\beta$) is drawn against the
 extinction rate $\beta$ and it is obvious that
 c($\beta$) = k ($\beta$ - ${\beta}_{\mbox{c}}$)
 with ${\beta}_{\mbox{c}}$ = 0.8362 and k $\approx$ 2.
 Whereas ${\beta}_{\mbox{c}}$ can be clearly identified as the critical
 extinction rate where the transition from freely moving to jammed traffic
 takes place the factor k is still an open question.\\
 If we pass over to the density profiles for $\beta$ $<$ ${\beta}_{c}$
 it can be easily seen in Figs 6c,d that the density profiles have
 the form $\rho$(i = 6+5n) =
 ${\rho}_{max}$($\beta$)[1 - ${\mbox{e}}^{ \mbox{c($\beta$) i}}$].\\
 \\
 The behaviour of the density profiles described in this section has the
 following physical explanation:\\
 As it is well-known, the right boundary has no effect on the density
 profiles for $\beta$ = 1. With decreasing $\beta$, however, there is
 a growing probability of a blockage at the end of the system, i. e., cars
 are increasingly hindered from moving out of the system. Consequently,
 a jam develops showing the growing influence of the right boundary with
 decreasing $\beta$. For $\beta$ $>$ ${\beta}_{c}$ the influence of the
 right boundary exponentially diminishes (Fig 6b).
 Fig 6a further shows that the left boundary conditions are still valid for
 the whole system, what can be seen at the oscillations of the
 density profile and in the constant value $\rho$ (i) = $\frac{1}{3}$ on the
 sites i = 4+5n and i = 5+5n characteristic for the case
 $\alpha$ = $\beta$ = 1.
 For decreasing $\beta$ the jam and with it the influence of the right boundary
 expands to the left.\\
 At $\beta$ = ${\beta}_{c}$ the repercussion of the right boundary reaches the
 left boundary, and the decay of the jam is proportional to i. Simultaneously,
 the influence of the left boundary is still present in the whole system, too,
 which manifests itself in the oscillations in the density profile going from
 the left to the right boundary (Fig 6a). So it can be said that for the
 extinction rate $\beta$ = ${\beta}_{c}$ the influence of the left and that
 of the right boundary coexist in the whole system.\\
 However, beginning from the right the oscillations vanish when the
 extinction rate is further decreased (Fig 6a). This indicates that the
 influence of the left boundary is pushed back for $\beta$ $<$ ${\beta}_{c}$.
 The form
 $\rho$(i = 6+5n) =
 ${\rho}_{max}$($\beta$)[1 - ${\mbox{e}}^{ \mbox{c($\beta$) i}}$]
 shows the decrease of unoccupied sites and may be a hint at a symmetry
 around the transition point.
 \\
 For very small $\beta$, the left boundary does not have any
 relevance at all for the movement of the cars in the bulk.\\
 \\
 Finally, let us say some words about the maximum value
 ${\rho}_{max}$($\beta$) = $max[ \rho \mbox{(i = 6+5n)}]$.
 From Fig 6a it is obvious that
 ${\rho}_{max}$($\beta$) can be identified with the density on site i = 1021,
 ${\rho}_{max}$($\beta$) = $\rho$ ($\beta$, i = 1021). From Fig 6e it
 turns out then that
 \begin{eqnarray}
 {\rho}_{max}( \beta ) = {\rho}_{max}({\beta}_{c}) \,\
           {e}^{{\mbox{k}}_{1} (\beta - {\beta}_{c})}
 \,\ \,\ \mbox{  for  } \beta > {\beta}_{c}\nonumber \\
 {\rho}_{max}( \beta ) = 1 + {\mbox{k}}_{2} \beta \hspace*{2cm}
 \,\ \,\ \mbox{  for  } \beta < {\beta}_{c}\nonumber
 \end{eqnarray}
 (${\mbox{k}}_{1}$ $\approx$ -24.46; ${\mbox{k}}_{2}$ $\approx$ -0.8;
 ${\beta}_{c}$ = 0.8362). Thus, the transition from freely moving to
 jammed traffic is reflected at the right boundary, too.\\

 \subsection{$\beta$ = 1-$\alpha$}
 \noindent
 We have already mentioned that for $\beta$ = 1-$\alpha$ we have rather
 similar conditions at the left and at the right boundary and therefore,
 systems with open and with periodic boundary conditions can be compared
 at best with each other in that case.\\
 We must keep in mind, however, that there are significant differences for
 $\beta$ = 1-$\alpha$, too, especially if the randomization probability
 is p = 0: In systems with periodic boundary conditions the movement of the
 cars is fully deterministic and the car density $\rho$ in the system
 keeps constant. Each site in the system has the same probability of being
 occupied and therefore, the density profiles of systems with periodic boundary
 conditions are constants with the value $\rho$ (the latter statement is
 also valid for randomization probabilities p $>$ 0).
 For corresponding systems with open boundary conditions - due to the
 injection rate $\alpha$ and the extinction rate $\beta$ - we always have
 a non-deterministic element at the boundaries of the system, also for the
 randomization probability p = 0 (which only refers to the movement in the
 bulk).\\
 \\
 Generally speaking, the density profiles for $\beta$ = 1 - $\alpha$ show a
 similar behaviour as those for the case $\alpha$ = 1: For very low extinction
 rates (and high injection rates) the density profiles are identical with the
 density profile of a corresponding system with periodic boundary
 conditions. For high $\beta$ (and low $\alpha$) the density profiles
 show the periodic structure already known from the previous sections as
 a typical feature of the free flow regime. At ${\beta}_{c}$ = 0.56 (and
 ${\alpha}_{c}$ = 0.44) the transition from free flow to jamming takes place.
 For $\beta$ $>$ ${\beta}_{c}$ the curves have the form
 $\rho$ (i = 6+5n) =
 ${\rho}_{max}$( $\beta$ ) ${\mbox{e}}^{ \mbox{c}( \beta ) \mbox{(i-L)}}$,
 for $\beta$ $<$ ${\beta}_{c}$
 $\rho$(i = 6+5n) =
 ${\rho}_{max}$($\beta$)[1 - ${\mbox{e}}^{ \mbox{c($\beta$) i}}$]
 and for $\beta$ = ${\beta}_{c}$ we have a straight line. The only difference
 to the $\alpha$ = 1 case is the value of the critical extinction rate and
 of k: For $\beta$ = 1-$\alpha$ we have ${\beta}_{c}$ = 0.56 and
 k = 3.75.\\
 \\
 In Section III we have already mentioned that in the high density regime
 the global density (current) for $\beta$ = 1-$\alpha$ is identical with the
 global density (current) for $\alpha$ = 1 and in the low density regime with
 the global density (current) for $\beta$ = 1.
 From Figs 8a,b it is obvious that similar effects can be also observed for
 the density profiles, too. Having a closer look at them we see that the
 profiles  for $\beta$ = 1-$\alpha$ and $\beta$ = 1 are identical if the
 injection rate $\alpha$ is low. For increasing $\alpha$ the density
 profiles for $\beta$ = 1-$\alpha$ start to lift at the end of the system
 indicating the growing influence of the right boundary on the system for
 increasing $\alpha$. On the other hand, comparing the density profiles
 for $\beta$ = 1-$\alpha$ and $\alpha$ = 1 with each other we see that they
 are identical for very low $\beta$. For increasing $\beta$ the density
 profile ''drops'' at the beginning of the system. Accordingly, this
 behaviour shows the growing influence of the left boundary on the system.
 In the transition regime, however, the density profiles for
 $\beta$ = 1-$\alpha$ are very different from those for the cases
 $\alpha$ = 1 and $\beta$ = 1.\\
 \section{Correlation Functions}
 \noindent
 In this section we consider correlation functions
 \begin{eqnarray}
 \mbox{C(i,t) = } < \eta \mbox{(i',t'), } \eta \mbox{(i'+i,t'+t)}
 {>}_{\mbox{\small{i',t'}}} - {\rho}^{2}
 \nonumber
 \end{eqnarray}
 for short ranges with
 \begin{eqnarray}
 && \eta \mbox{(i',t') = 1 \,\ \,\   if site i' is occupied at time t'}
 \nonumber \\
 && \eta \mbox{(i',t') = 0 \,\ \,\  else}
 \nonumber
 \end{eqnarray}
 This kind of correlation functions has been already investigated for systems
 with periodic boundary conditions and the randomization probability p = 0.5
 in \cite {Chey:Kert:Schr}. It turned out that in the free flow regime there
 is a propagating peak at i = ${\mbox{v}}_{max}$t with a
 shoulder at i = ${\mbox{v}}_{max}$t - 1 and with
 anticorrelations around it. The density where these anticorrelations are
 maximally developed is defined as the density where the transition from free
 flow to jamming takes place. For higher densities a jamming peak occurs at
 i = -1 \cite {Chey:Kert:Schr}.\\
 \\
 It would be interesting to see if these features can be also found for systems
 with open boundary conditions. But considering the deterministic case in this
 paper we should investigate the correlation functions for systems with
 periodic boundaries and p = 0 first. From Fig 8a it can be seen that the
 propagating peak is sharp and that there are further peaks at
 i = ${\mbox{v}}_{max}$t $\pm$ 6n (n = 1, 2,...) as the
 movement of the cars in the ring is deterministic. Due to the fact that
 the initial configuration is random, however, these peaks become smaller
 and smaller with increasing n. Between the peaks anticorrelations are
 observed which are best developed around the peak at
 i = ${\mbox{v}}_{max}$t. Generally speaking it can be said
 that in the free flow regime the correlation functions C(i,t) are
 symmetric around the site i = ${\mbox{v}}_{max}$t.
 \\
 Coming from low densities the anticorrelations become deeper and deeper with increasing $\rho$. At
 $\rho$ = ${\rho}_{c}$ = $\frac{1}{{\mbox{v}_{max}}+1}$
 the car distribution is well-defined: all vehicles drive with the maximum
 velocity ${\mbox{v}}_{max}$ = 5 and between two neighbouring
 cars there are ${\mbox{v}}_{max}$ = 5 empty sites each.
 Correspondingly, the correlation function for $\rho$ = ${\rho}_{c}$ is
 periodic with
 \begin{eqnarray}
 && \mbox{C(i,t) = $\rho$ -${\rho}^{2}$ \,\ \,\ if
 i = ${\mbox{v}}_{max}$t $\pm$ 6n}
 \nonumber
 \\
 && \mbox{C(i,t) = - ${\rho}^{2}$ \,\ \,\ else}
 \nonumber
 \end{eqnarray}
 At this density where the transition from free flow to jamming takes place
 the anticorrelations reach their minimum.\\
 For higher densities a jamming peak develops at i = -1 (due to the hindrance
 the back car feels in the jam) with anticorrelations at i = $\pm$ 1. At all
 other sites peaks and anticorrelations vanish. If the density further
 increases fewer and fewer cars move (with v $>$ 0) and therefore, the
 anticorrelations at i = $\pm$ 1 disappear. Corresponding to the symmetry
 around i = ${\mbox{v}}_{max}$t in the free flow regime, the
 correlation functions for $\rho > {\rho}_{c}$ are symmetric around i = -1.\\
 \\
 Let us turn back to systems with open boundary conditions which is the
 real topic of this paper. In Fig 8b we consider correlation functions
 from the middle of the system because the influence of the boundaries is
 minimal there. It is obvious that for high densities the correlation
 functions in systems with open boundary conditions are nearly identical.
 Merely the minor maxima at i = $\pm$ 2 in Fig 8a shift onto i = $\pm$ 3 in
 Fig 8b.
 \\
 If the density in the system is low, however, the situation is completely
 different: For systems with open boundaries we have a random element at the
 boundaries where cars are randomly created and deleted at each time step.
 Therefore, due to the permanent presence of randomization even if the
 movement in the bulk is deterministic we can only observe the propagating
 at i = ${\mbox{v}}_{max}$t (and a very small one at
 i = ${\mbox{v}}_{max}$t $\pm$ 6). Around the propagating
 peak there are anticorrelations, too, but they are not so well-developed as
 the anticorrelations of corresponding correlation function in the case of
 periodic boundary conditions. However, a common feature of systems with
 open and periodic boundary conditions is the symmetry of the correlation
 functions around i = ${\mbox{v}}_{max}$t in the free flow
 regime.\\
 \\
 As we have already mentioned the anticorrelations around the propagating
 peak play an important role in systems with periodic boundary conditions
 and we will now discuss the question if similar features can be observed
 for systems with open boundary conditions.
 In Figs 8b-d we consider short-range correlation functions at the beginning,
 the middle, and the end of the system. Coming from high extinction rates
 $\beta$ (with $\beta$ = 1-$\alpha$) the anticorrelations become deeper and
 deeper everywhere in the system. But what it is interesting is the fact that
 the anticorrelations start to vanish again at an extinction rate where the
 influence of the right boundary reaches the corresponding sites (see also
 Section V.C). Strictly speaking the anticorrelations start to vanish at about
 $\beta$ = 0.42 (at the end), $\beta$ = 0.43 (in the middle), and
 $\beta$ = 0.44 (at the beginning).\\
 Therefore to sum it up it can be said that the injection rate (extinction
 rate) where the probability to find a car in the neighbourhood of another
 car is minimal (that means, where the anticorrelations start to vanish) can be
 considered  as the injection rate ${\alpha}_{c}$ (extinction rate
 ${\beta}_{c}$) where the transition from free flow to jamming takes place.\\
 \section{Conclusions and Discussion}
 \noindent
 Systems with open boundaries where cars move deterministically with
 maximum velocity ${\mbox{v}}_{max}$ $>$ 1 show interesting features mainly
 resulting from the competition of the left and of the right boundary for
 the influence in the system and from the existence of so-called ''buffers''.
 \\
 \\
 The latter plays a fundamental role at the comparison of systems with
  ${\mbox{v}}_{max}$ $\ge$ 3 and ${\mbox{v}}_{max}$ = 1. One of the most
 important questions in this context is why the border between free flow
 and jamming for ${\mbox{v}}_{max}$ $\ge$ 3 has such a different course
 than the corresponding border for the case ${\mbox{v}}_{max}$ = 1.
 By simple
 analytical considerations it turns out that - as a consequence of the
 hindrance an injected car feels from the front car - spaces $>$
 ${\mbox{v}}_{max}$ develop for high injection rates $\alpha$ (for the special
 case of $\alpha$ = $\beta$ = 1 there are alternately
 2(${\mbox{v}}_{max}$-1) and ${\mbox{v}}_{max}$ sites between
 neighbouring cars for all ${\mbox{v}}_{max}$ $>$ 1). That means in addition
 to the expected ${\mbox{v}}_{max}$ sites further sites occur which are the
 reason why the maximum current is found at
 $\alpha$ $<$ 1 and $\frac{5}{6}$ $\le$ $\beta$ $\le$ 1
 for ${\mbox{v}}_{max}$ $\ge$ 5. We call these additional sites ''buffers''
 because they also have a buffer effect at the
 end of the system: Due to the buffers the development of jamming waves is
 suppressed up to an injection rate $\beta$ = $\frac{5}{6}$ (for high $\alpha$
 and
 ${\mbox{v}}_{max}$ $\ge$ 5)
 and this buffer effect is responsible for the characteristic course of the
 free flow - jamming border for ${\mbox{v}}_{max}$ $\ge$ 5.
 The transition from the free flow to the congested phase is of first order
 and accompagnied by the collapse of the buffers.
 \\
 In this context, it should be emphasized that the occurrence of buffers -
 and consequently the specific features of the ${\mbox{v}}_{max}$ $\ge$ 3
 model - is due to the parallel updating mechanism and not an effect of the
 particular injection rule.
 Naturally, there are other possibilities of generalizing the ASEP to
 ${\mbox{v}}_{max}$ $>$ 1, for example, one could keep the existence of the
 car at i = 0 if i = 1 is occupied by another car. Simulations based on this
 alternative rule show that the phase diagram and the
 $\alpha$,$\beta$-dependence of the current are
 qualitatively the same as the corresponding quantities considered in this
 paper. This has been confirmed by analytical investigations of the special
 case ${\mbox{v}}_{max}$ = 5,
 $\alpha$ = $\beta$ = 1 (according to Section III) where buffers occur, too.
 \\
 As global density and current (from now on we exclusively refer again to the
 injection rule defined in Section II) show no qualitative differences for
 ${\mbox{v}}_{max}$ $\ge$ 5, a detailed analysis of the influence of the
 boundary conditions on the system (by means of density profiles and
 short-range correlation functions) is confined to the maximum velocity
 ${\mbox{v}}_{max}$ $\ge$ 5.
 Furthermore, our numerical investigations are based on systems with size
 L = 1024. It must be mentioned here that finite size effects occur which do
 not play an important role, however: the discontinuous transition in the
 global density becomes continuous and the value for the critical extinction
 rate (${\beta}_{c}$ = 0.836 for $\alpha$ = 1 and L = 1024) is slightly
 higher than in the case L $\rightarrow$ $\infty$.
 \\
 \\
 If we consider a single site of the system
 there are three possibilities:
 \begin{eqnarray}
 && \mbox{A. the site is exclusively under the influence of the left
 boundary}
 \nonumber \\
 && \mbox{B. the site is exclusively under the influence of the right
 boundary}
 \nonumber \\
 && \mbox{C. the site is under the influence of both the left and the right
 boundary}
 \nonumber
 \end{eqnarray}
 In the free flow regime the system consists of sites of the types A and C,
 in the jamming regime of sites of the type B and C. The critical injection
 rate ${\alpha}_{c}$ (the critical extinction rate ${\beta}_{c}$) where the
 transition from freely moving to jammed traffic takes place, is the only
 $\alpha$ ($\beta$) where {\it all} sites of the system belong to the C-type,
 that means, where the influence of the left and of the right boundary coexists
 in the whole system. The farther we go away from the transition point the
 stronger is the dominance of the A-sites in the free flow regime and of the
 B-sites in the jamming region.\\
 In the free flow (jamming) regime the current does not depend on
 the injection rate $\alpha$ (extinction rate $\beta$) what confirms the
 dominance of the left (right) boundary influence in the free flow
 (jamming) regime.\\
 \\
 Comparing the density profiles for ${\mbox{v}}_{max}$ = 5 with those for
 ${\mbox{v}}_{max}$ = 1 we see that the most significant differences are
 found in the free flow regime: In the free flow regime the density
 profile for ${\mbox{v}}_{max}$ = 5 shows periodic structure with the period
 $\Delta$i = ${\mbox{v}}_{max}$ = 5 which is due to the free movement of the
 cars. Another interesting result is the fact that the sites i = 6+5n
 (n = 1, 2, ...) are never occupied when they are beyond the sphere of
 influence of the right boundary. In the jamming regime, however, the density
 profiles for ${\mbox{v}}_{max}$ = 5 and ${\mbox{v}}_{max}$ = 1 are nearly
 the same. The investigation of the density profiles, especially the behaviour
 on the sites  = 6 + 5n enables the precise localization of the phase
 transition.\\
 \\
 The short-range correlation functions C(i,t) show that
 there are parallels between systems with open and with periodic boundary
 conditions, which are the following: In the free flow regime C(i,t) is
 symmetric around the site i = ${\mbox{v}}_{max}$t and in the jamming
 region around i = -1 which may be a hint at a symmetry.
 Free flow (jamming) is characterized by a peak at i = ${\mbox{v}}_{max}$t
 (i = -1) with anticorrelations around it. Furthermore, in systems with open
 or periodic boundary conditions the anticorrelations around the free flow peak
 are maximally developed when the transition from free flow to jamming
 takes place.\\
 \section{Acknowledgments}
 This work was supported by the Land of North Rhine-Westphalia and by the
 OTKA(T029985).
 \newpage

 \section{
APPENDIX A: Disturbance at the beginning of the system 
}

\noindent
$\hspace*{-1cm}$(a) 
$\hspace*{1cm} \vdots$ $\hspace*{10cm} \vdots$ \\
\normalsize{
(2) ${x}^{2}$(3)${x}^{5}$ $\cdots$ 
(\^v)${x}^{2(\mbox{\^v}-1)}$
(\^v)${x}^{\mbox{\^v}}$
(\^v)${x}^{2(\mbox{\^v}-1)}$
(\^v)${x}^{\mbox{\^v}}$
(\^v)${x}^{2(\mbox{\^v}-1)}$
(\^v)
$\cdots$ \\
(1) ${x}^{1}$(3)${x}^{3}$ $\cdots$ 
(\^v)${x}^{2(\mbox{\^v}-1)}$
(\^v)${x}^{\mbox{\^v}}$
(\^v)${x}^{2(\mbox{\^v}-1)}$
(\^v)${x}^{\mbox{\^v}}$
(\^v)${x}^{2(\mbox{\^v}-1)}$
(\^v)
$\cdots$ \\
(0) (2)${x}^{3}$(4) $\cdots$ 
(\^v)${x}^{2(\mbox{\^v}-1)}$
(\^v)${x}^{\mbox{\^v}}$
(\^v)${x}^{2(\mbox{\^v}-1)}$
(\^v)${x}^{\mbox{\^v}}$
(\^v)${x}^{2(\mbox{\^v}-1)}$
(\^v)
$\cdots$ \\
(-) ${x}^{2}$(3)${x}^{5}$ $\cdots$ 
(\^v)${x}^{2(\mbox{\^v}-1)}$
(\^v)${x}^{\mbox{\^v}}$
(\^v)${x}^{2(\mbox{\^v}-1)}$
(\^v)${x}^{\mbox{\^v}}$
(\^v)${x}^{2(\mbox{\^v}-1)}$
(\^v)
$\cdots$ $\leftarrow$ no car \\
(5) ${x}^{5}$(4)${x}^{7}$ $\cdots$ 
(\^v)${x}^{2(\mbox{\^v}-1)}$
(\^v)${x}^{\mbox{\^v}}$
(\^v)${x}^{2(\mbox{\^v}-1)}$
(\^v)${x}^{\mbox{\^v}}$
(\^v)${x}^{2(\mbox{\^v}-1)}$
(\^v)
$\cdots$ \hspace{0.3cm} injected! \\
(4) ${x}^{4}$(4)${x}^{4}$ $\cdots$ 
(\^v)${x}^{2(\mbox{\^v}-1)}$
(\^v)${x}^{\mbox{\^v}}$
(\^v)${x}^{2(\mbox{\^v}-1)}$
(\^v)${x}^{\mbox{\^v}}$
(\^v)${x}^{2(\mbox{\^v}-1)}$
(\^v)
$\cdots$ \\
(3) ${x}^{3}$(4)${x}^{4}$ $\cdots$ 
(\^v)${x}^{2(\mbox{\^v}-1)}$
(\^v)${x}^{\mbox{\^v}}$
(\^v)${x}^{2(\mbox{\^v}-1)}$
(\^v)${x}^{\mbox{\^v}}$
(\^v)${x}^{2(\mbox{\^v}-1)}$
(\^v)
$\cdots$ \\
(2) ${x}^{2}$(4)${x}^{4}$ $\cdots$ 
(\^v)${x}^{2(\mbox{\^v}-1)}$
(\^v)${x}^{\mbox{\^v}}$
(\^v)${x}^{2(\mbox{\^v}-1)}$
(\^v)${x}^{\mbox{\^v}}$
(\^v)${x}^{2(\mbox{\^v}-1)}$
(\^v)
$\cdots$ \\
(1) ${x}^{1}$(3)${x}^{4}$ $\cdots$ 
(\^v)${x}^{2(\mbox{\^v}-1)}$
(\^v)${x}^{\mbox{\^v}}$
(\^v)${x}^{2(\mbox{\^v}-1)}$
(\^v)${x}^{\mbox{\^v}}$
(\^v)${x}^{2(\mbox{\^v}-1)}$
(\^v)
$\cdots$ \\
(0) (2)${x}^{3}$(4) $\cdots$ 
(\^v)${x}^{2(\mbox{\^v}-1)}$
(\^v)${x}^{\mbox{\^v}}$
(\^v)${x}^{2(\mbox{\^v}-1)}$
(\^v)${x}^{\mbox{\^v}}$
(\^v)${x}^{2(\mbox{\^v}-1)}$
(\^v)
$\cdots$ \\
(2) ${x}^{2}$(3)${x}^{5}$ $\cdots$ 
(\^v)${x}^{2(\mbox{\^v}-1)}$
(\^v)${x}^{\mbox{\^v}}$
(\^v)${x}^{2(\mbox{\^v}-1)}$
(\^v)${x}^{\mbox{\^v}}$
(\^v)${x}^{2(\mbox{\^v}-1)}$
(\^v)
$\cdots$ \\
(1) ${x}^{1}$(3)${x}^{3}$ $\cdots$ 
(\^v)${x}^{2(\mbox{\^v}-1)}$
(\^v)${x}^{\mbox{\^v}}$
(\^v)${x}^{2(\mbox{\^v}-1)}$
(\^v)${x}^{\mbox{\^v}}$
(\^v)${x}^{2(\mbox{\^v}-1)}$
(\^v)
$\cdots$ \\
(0) (2)${x}^{3}$(4) $\cdots$ 
(\^v)${x}^{2(\mbox{\^v}-1)}$
(\^v)${x}^{\mbox{\^v}}$
(\^v)${x}^{2(\mbox{\^v}-1)}$
(\^v)${x}^{\mbox{\^v}}$
(\^v)${x}^{2(\mbox{\^v}-1)}$
(\^v)
$\cdots$ \\
$\hspace*{1cm} \vdots$ $\hspace*{10cm} \vdots$ \\
(2) ${x}^{2}$(3)${x}^{5}$ $\cdots$ 
(\^v)${x}^{\mbox{\^v}}$
(\^v)${x}^{2(\mbox{\^v}-1)}$
(\^v)${x}^{2\mbox{\^v}-3}$
(\^v)${x}^{\mbox{\^v}}$
(\^v)${x}^{\mbox{\^v}}$
(\^v)${x}^{\mbox{\^v}}$
(\^v)
$\cdots$ $\cdots$ $\cdots$ \\
(1) ${x}^{1}$(3)${x}^{3}$ $\cdots$ $\cdots$ 
(\^v)${x}^{\mbox{\^v}}$
(\^v)${x}^{2(\mbox{\^v}-1)}$
(\^v)${x}^{2\mbox{\^v}-3}$
(\^v)${x}^{\mbox{\^v}}$
(\^v)${x}^{\mbox{\^v}}$
(\^v)${x}^{\mbox{\^v}}$
(\^v)
$\cdots$ $\cdots$ \\
(0) (2)${x}^{3}$(4) $\cdots$ $\cdots$ $\cdots$ 
(\^v)${x}^{\mbox{\^v}}$
(\^v)${x}^{2(\mbox{\^v}-1)}$
$\underbrace{
(\mbox{\^v}){x}^{2\mbox{\^v}-3}
(\mbox{\^v}){x}^{\mbox{\^v}}
(\mbox{\^v}){x}^{\mbox{\^v}}
}$
(\^v)${x}^{\mbox{\^v}}$
(\^v)
$\cdots$ \\
\hspace*{7cm} effect of disturbance \\
}
\vspace*{0.5cm}

\noindent
$\hspace*{-1cm}$ {\normalsize(b)} 
$\hspace*{1cm} \vdots$ $\hspace*{10cm} \vdots$ \\
\normalsize{
(1) ${x}^{1}$(3)${x}^{3}$ $\cdots$ 
(\^v)${x}^{2(\mbox{\^v}-1)}$
(\^v)${x}^{\mbox{\^v}}$
(\^v)${x}^{2(\mbox{\^v}-1)}$
(\^v)${x}^{\mbox{\^v}}$
(\^v)${x}^{2(\mbox{\^v}-1)}$
(\^v)
$\cdots$ \\
(0) (2)${x}^{3}$(4) $\cdots$ 
(\^v)${x}^{2(\mbox{\^v}-1)}$
(\^v)${x}^{\mbox{\^v}}$
(\^v)${x}^{2(\mbox{\^v}-1)}$
(\^v)${x}^{\mbox{\^v}}$
(\^v)${x}^{2(\mbox{\^v}-1)}$
(\^v)
$\cdots$ \\
(2) ${x}^{2}$(3)${x}^{5}$ $\cdots$ 
(\^v)${x}^{2(\mbox{\^v}-1)}$
(\^v)${x}^{\mbox{\^v}}$
(\^v)${x}^{2(\mbox{\^v}-1)}$
(\^v)${x}^{\mbox{\^v}}$
(\^v)${x}^{2(\mbox{\^v}-1)}$
(\^v)
$\cdots$ \\
(-) ${x}^{1}$(3)${x}^{3}$ $\cdots$ 
(\^v)${x}^{2(\mbox{\^v}-1)}$
(\^v)${x}^{\mbox{\^v}}$
(\^v)${x}^{2(\mbox{\^v}-1)}$
(\^v)${x}^{\mbox{\^v}}$
(\^v)${x}^{2(\mbox{\^v}-1)}$
(\^v)
$\cdots$ $\leftarrow$ no car \\
(4) ${x}^{4}$(4)${x}^{4}$ $\cdots$ 
(\^v)${x}^{2(\mbox{\^v}-1)}$
(\^v)${x}^{\mbox{\^v}}$
(\^v)${x}^{2(\mbox{\^v}-1)}$
(\^v)${x}^{\mbox{\^v}}$
(\^v)${x}^{2(\mbox{\^v}-1)}$
(\^v)
$\cdots$ \hspace{0.3cm} injected! \\
(3) ${x}^{3}$(4)${x}^{4}$ $\cdots$ 
(\^v)${x}^{2(\mbox{\^v}-1)}$
(\^v)${x}^{\mbox{\^v}}$
(\^v)${x}^{2(\mbox{\^v}-1)}$
(\^v)${x}^{\mbox{\^v}}$
(\^v)${x}^{2(\mbox{\^v}-1)}$
(\^v)
$\cdots$ \\
(2) ${x}^{2}$(4)${x}^{4}$ $\cdots$ 
(\^v)${x}^{2(\mbox{\^v}-1)}$
(\^v)${x}^{\mbox{\^v}}$
(\^v)${x}^{2(\mbox{\^v}-1)}$
(\^v)${x}^{\mbox{\^v}}$
(\^v)${x}^{2(\mbox{\^v}-1)}$
(\^v)
$\cdots$ \\
(1) ${x}^{1}$(3)${x}^{4}$ $\cdots$ 
(\^v)${x}^{2(\mbox{\^v}-1)}$
(\^v)${x}^{\mbox{\^v}}$
(\^v)${x}^{2(\mbox{\^v}-1)}$
(\^v)${x}^{\mbox{\^v}}$
(\^v)${x}^{2(\mbox{\^v}-1)}$
(\^v)
$\cdots$ \\
(0) (2)${x}^{3}$(4) $\cdots$ 
(\^v)${x}^{2(\mbox{\^v}-1)}$
(\^v)${x}^{\mbox{\^v}}$
(\^v)${x}^{2(\mbox{\^v}-1)}$
(\^v)${x}^{\mbox{\^v}}$
(\^v)${x}^{2(\mbox{\^v}-1)}$
(\^v)
$\cdots$ \\
(2) ${x}^{2}$(3)${x}^{5}$ $\cdots$ 
(\^v)${x}^{2(\mbox{\^v}-1)}$
(\^v)${x}^{\mbox{\^v}}$
(\^v)${x}^{2(\mbox{\^v}-1)}$
(\^v)${x}^{\mbox{\^v}}$
(\^v)${x}^{2(\mbox{\^v}-1)}$
(\^v)
$\cdots$ \\
(1) ${x}^{1}$(3)${x}^{3}$ $\cdots$ 
(\^v)${x}^{2(\mbox{\^v}-1)}$
(\^v)${x}^{\mbox{\^v}}$
(\^v)${x}^{2(\mbox{\^v}-1)}$
(\^v)${x}^{\mbox{\^v}}$
(\^v)${x}^{2(\mbox{\^v}-1)}$
(\^v)
$\cdots$ \\
(0) (2)${x}^{3}$(4) $\cdots$ 
(\^v)${x}^{2(\mbox{\^v}-1)}$
(\^v)${x}^{\mbox{\^v}}$
(\^v)${x}^{2(\mbox{\^v}-1)}$
(\^v)${x}^{\mbox{\^v}}$
(\^v)${x}^{2(\mbox{\^v}-1)}$
(\^v)
$\cdots$ \\
$\hspace*{1cm} \vdots$ $\hspace*{10cm} \vdots$ \\
(2) ${x}^{2}$(3)${x}^{5}$ $\cdots$ 
(\^v)${x}^{\mbox{\^v}}$
(\^v)${x}^{2(\mbox{\^v}-1)}$
(\^v)${x}^{2\mbox{\^v}-3}$
(\^v)${x}^{\mbox{\^v}}$
(\^v)${x}^{\mbox{\^v}}$
(\^v)${x}^{\mbox{\^v}}$
(\^v)
$\cdots$ $\cdots$ $\cdots$\\
(1) ${x}^{1}$(3)${x}^{3}$ $\cdots$ $\cdots$ 
(\^v)${x}^{\mbox{\^v}}$
(\^v)${x}^{2(\mbox{\^v}-1)}$
(\^v)${x}^{2\mbox{\^v}-3}$
(\^v)${x}^{\mbox{\^v}}$
(\^v)${x}^{\mbox{\^v}}$
(\^v)${x}^{\mbox{\^v}}$
(\^v)
$\cdots$ $\cdots$ \\
(0) (2)${x}^{3}$(4) $\cdots$ $\cdots$ $\cdots$ 
(\^v)${x}^{\mbox{\^v}}$
(\^v)${x}^{2(\mbox{\^v}-1)}$
$\underbrace{
(\mbox{\^v}){x}^{2\mbox{\^v}-3}
(\mbox{\^v}){x}^{\mbox{\^v}}
(\mbox{\^v}){x}^{\mbox{\^v}}
}$
(\^v)${x}^{\mbox{\^v}}$
(\^v)
$\cdots$ \\
\hspace*{7cm} effect of disturbance \\
}
\vspace*{0.5cm}

\noindent
$\hspace*{-1cm}$ {\normalsize (c)} 
$\hspace*{1cm} \vdots$ $\hspace*{10cm} \vdots$ \\
\normalsize{
(0) (2)${x}^{3}$(4) $\cdots$ 
(\^v)${x}^{2(\mbox{\^v}-1)}$
(\^v)${x}^{\mbox{\^v}}$
(\^v)${x}^{2(\mbox{\^v}-1)}$
(\^v)${x}^{\mbox{\^v}}$
(\^v)${x}^{2(\mbox{\^v}-1)}$
(\^v)
$\cdots$ \\
(2) ${x}^{2}$(3)${x}^{5}$ $\cdots$ 
(\^v)${x}^{2(\mbox{\^v}-1)}$
(\^v)${x}^{\mbox{\^v}}$
(\^v)${x}^{2(\mbox{\^v}-1)}$
(\^v)${x}^{\mbox{\^v}}$
(\^v)${x}^{2(\mbox{\^v}-1)}$
(\^v)
$\cdots$ \\
(1) ${x}^{1}$(3)${x}^{3}$ $\cdots$ 
(\^v)${x}^{2(\mbox{\^v}-1)}$
(\^v)${x}^{\mbox{\^v}}$
(\^v)${x}^{2(\mbox{\^v}-1)}$
(\^v)${x}^{\mbox{\^v}}$
(\^v)${x}^{2(\mbox{\^v}-1)}$
(\^v)
$\cdots$ \\
(-) (2)${x}^{3}$(4) $\cdots$ 
(\^v)${x}^{2(\mbox{\^v}-1)}$
(\^v)${x}^{\mbox{\^v}}$
(\^v)${x}^{2(\mbox{\^v}-1)}$
(\^v)${x}^{\mbox{\^v}}$
(\^v)${x}^{2(\mbox{\^v}-1)}$
(\^v)
$\cdots$ $\leftarrow$ no car \\
(2) ${x}^{2}$(3)${x}^{5}$ $\cdots$ 
(\^v)${x}^{2(\mbox{\^v}-1)}$
(\^v)${x}^{\mbox{\^v}}$
(\^v)${x}^{2(\mbox{\^v}-1)}$
(\^v)${x}^{\mbox{\^v}}$
(\^v)${x}^{2(\mbox{\^v}-1)}$
(\^v)
$\cdots$ injected! \\
(1) ${x}^{1}$(3)${x}^{3}$ $\cdots$  
(\^v)${x}^{2(\mbox{\^v}-1)}$
(\^v)${x}^{\mbox{\^v}}$
(\^v)${x}^{2(\mbox{\^v}-1)}$
(\^v)${x}^{\mbox{\^v}}$
(\^v)${x}^{2(\mbox{\^v}-1)}$
(\^v)
$\cdots$ \\
(0) (2)${x}^{3}$(4) $\cdots$  
(\^v)${x}^{2(\mbox{\^v}-1)}$
(\^v)${x}^{\mbox{\^v}}$
(\^v)${x}^{2(\mbox{\^v}-1)}$
(\^v)${x}^{\mbox{\^v}}$
(\^v)${x}^{2(\mbox{\^v}-1)}$
(\^v)
$\cdots$ \\
\\
$\hspace*{4.5cm}$ $\Rightarrow$ no effect of disturbance 
}
\\

 \section{
APPENDIX B: Disturbance at the end of the system}
\noindent
$\hspace*{-1cm}$(a) 
$\hspace*{2cm} \vdots$ $\hspace*{4cm} \vdots$ \\
\normalsize{
$\cdots$ 
${x}^{2(\mbox{\^v}-1)}$ \hspace{0.2cm}
(\^v) \hspace{0.6cm}
${x}^{\mbox{\^v}}$ \hspace{0.4cm}
(\^v) \hspace{0.4cm} 
${x}^{2(\mbox{\^v}-1)}$ \hspace{0.2cm}
(\^v)\\  
$\cdots$
$\, \,$
(\^v) \hspace{0.4cm}
${x}^{2(\mbox{\^v}-1)}$ \hspace{0.1cm}
(\^v) \hspace{0.5cm}
${x}^{\mbox{\^v}}$ \hspace{0.6cm}
(\^v) \hspace{0.4cm}
${x}^{\mbox{\^v}-1}$\\
$\cdots$
$\, \,$ 
(\^v) \hspace{0.6cm}
${x}^{\mbox{\^v}}$ \hspace{0.6cm}
(\^v) \hspace{0.2cm}
${x}^{2(\mbox{\^v}-1)}$ \hspace{0.2cm}
(\^v) \hspace{0.6cm}
${x}^{\mbox{\^v}}$\\
$\cdots$ 
${x}^{2(\mbox{\^v}-1)}$ \hspace{0.2cm}
(\^v) \hspace{0.6cm}
${x}^{\mbox{\^v}}$ \hspace{0.4cm}
(\^v) \hspace{0.4cm} 
${x}^{2(\mbox{\^v}-1)}$ \hspace{0.2cm}
(0) 
\hspace{0.2cm} $\leftarrow$ blockage! 
\\
$\cdots$ 
${x}^{2(\mbox{\^v}-1)}$ \hspace{0.2cm}
(\^v) \hspace{0.6cm}
${x}^{\mbox{\^v}}$ \hspace{0.3cm}
(\^v-2) \hspace{0.4cm} 
${x}^{\mbox{\^v}-2}$ \hspace{0.4cm}
(1)   
\\
$\cdots$
$\, \,$
(\^v) \hspace{0.4cm}
${x}^{2(\mbox{\^v}-1)}$ \hspace{0.1cm}
(\^v-2) \hspace{0.2cm}
${x}^{\mbox{\^v}-2}$ \hspace{0.1cm}
(\^v-1) \hspace{0.5cm}
${x}^{1}$\\
$\cdots$
$\, \,$ 
(\^v) \hspace{0.6cm}
${x}^{\mbox{\^v}}$ \hspace{0.6cm}
(\^v) \hspace{0.2cm}
${x}^{2(\mbox{\^v}-2)}$ \hspace{0.1cm}
(\^v-1) \hspace{0.5cm}
${x}^{2}$
$\, \, \, \,$
$\_\_\_\_\_\_\_\_\_\_\_\_\_\_\_\_\_\_\_\_\_$ \\
$\cdots$
$\, \,$
(\^v) \hspace{0.4cm}
${x}^{2(\mbox{\^v}-1)}$ \hspace{0.1cm}
(\^v) \hspace{0.5cm}
${x}^{\mbox{\^v}}$ \hspace{0.6cm}
(\^v) \hspace{0.4cm}
${x}^{\mbox{\^v}-1}$ 
$\, \,$ from here on \\
$\cdots$
$\, \,$ 
(\^v) \hspace{0.6cm}
${x}^{\mbox{\^v}}$ \hspace{0.6cm}
(\^v) \hspace{0.2cm}
${x}^{2(\mbox{\^v}-1)}$ \hspace{0.2cm}
(\^v) \hspace{0.6cm}
${x}^{\mbox{\^v}}$ 
$\, \, \,$ nothing reminds \\
$\cdots$ 
${x}^{2(\mbox{\^v}-1)}$ \hspace{0.2cm}
(\^v) \hspace{0.6cm}
${x}^{\mbox{\^v}}$ \hspace{0.4cm}
(\^v) \hspace{0.4cm} 
${x}^{2(\mbox{\^v}-1)}$ \hspace{0.2cm}
(\^v) 
$\, \, \,$ of the disturbance \\ 
}
\vspace*{0.5cm}

\noindent
$\hspace*{-1cm}$ {\normalsize(b)} 
$\hspace*{2cm} \vdots$ $\hspace*{4cm} \vdots$ \\
\normalsize{
$\cdots$
$\, \,$
(\^v) \hspace{0.4cm}
${x}^{2(\mbox{\^v}-1)}$ \hspace{0.1cm}
(\^v) \hspace{0.5cm}
${x}^{\mbox{\^v}}$ \hspace{0.6cm}
(\^v) \hspace{0.4cm}
${x}^{\mbox{\^v}-1}$\\
$\cdots$
$\, \,$ 
(\^v) \hspace{0.6cm}
${x}^{\mbox{\^v}}$ \hspace{0.6cm}
(\^v) \hspace{0.2cm}
${x}^{2(\mbox{\^v}-1)}$ \hspace{0.2cm}
(\^v) \hspace{0.6cm}
${x}^{\mbox{\^v}}$\\
$\cdots$ 
${x}^{2(\mbox{\^v}-1)}$ \hspace{0.2cm}
(\^v) \hspace{0.6cm}
${x}^{\mbox{\^v}}$ \hspace{0.4cm}
(\^v) \hspace{0.4cm} 
${x}^{2(\mbox{\^v}-1)}$ \hspace{0.2cm}
(\^v)\\  
$\cdots$
$\, \,$
(\^v) \hspace{0.4cm}
${x}^{2(\mbox{\^v}-1)}$ \hspace{0.1cm}
(\^v) \hspace{0.5cm}
${x}^{\mbox{\^v}}$ \hspace{0.5cm}
(\^v-1) \hspace{0.4cm}
${x}^{\mbox{\^v}-1}$ 
\hspace{0.2cm} $\leftarrow$ blockage! \\
$\cdots$
$\, \,$
${x}^{\mbox{\^v}}$ \hspace{0.6cm}
(\^v) \hspace{0.4cm}
${x}^{2(\mbox{\^v}-1)}$ 
(\^v-1) \hspace{0.4cm}
${x}^{\mbox{\^v}-1}$ \hspace{0.3cm}
(\^v) \\
$\cdots$
$\, \,$ 
(\^v) \hspace{0.6cm}
${x}^{\mbox{\^v}}$ \hspace{0.6cm}
(\^v) \hspace{0.2cm}
${x}^{2\mbox{\^v}-3}$ \hspace{0.4cm}
(\^v) \hspace{0.7cm}
${x}^{1}$ 
$\, \, \, \,$
$\_\_\_\_\_\_\_\_\_\_\_\_\_\_\_\_\_\_\_\_\_$ \\
$\cdots$
$\, \,$
(\^v) \hspace{0.4cm}
${x}^{2(\mbox{\^v}-1)}$ \hspace{0.1cm}
(\^v) \hspace{0.5cm}
${x}^{\mbox{\^v}}$ \hspace{0.6cm}
(\^v) \hspace{0.4cm}
${x}^{\mbox{\^v}-1}$ 
$\, \,$ from here on \\
$\cdots$
$\, \,$ 
(\^v) \hspace{0.6cm}
${x}^{\mbox{\^v}}$ \hspace{0.6cm}
(\^v) \hspace{0.2cm}
${x}^{2(\mbox{\^v}-1)}$ \hspace{0.2cm}
(\^v) \hspace{0.6cm}
${x}^{\mbox{\^v}}$ 
$\, \, \,$ nothing reminds \\
$\cdots$ 
${x}^{2(\mbox{\^v}-1)}$ \hspace{0.2cm}
(\^v) \hspace{0.6cm}
${x}^{\mbox{\^v}}$ \hspace{0.4cm}
(\^v) \hspace{0.4cm} 
${x}^{2(\mbox{\^v}-1)}$ \hspace{0.2cm}
(\^v) 
$\, \, \,$ of the disturbance \\ 
}
\vspace*{0.5cm}

\noindent
$\hspace*{-1cm}$ {\normalsize(c)} 
$\hspace*{2cm} \vdots$ $\hspace*{4cm} \vdots$ \\
\normalsize{
$\cdots$
$\, \,$ 
(\^v) \hspace{0.6cm}
${x}^{\mbox{\^v}}$ \hspace{0.6cm}
(\^v) \hspace{0.2cm}
${x}^{2(\mbox{\^v}-1)}$ \hspace{0.2cm}
(\^v) \hspace{0.6cm}
${x}^{\mbox{\^v}}$\\
$\cdots$ 
${x}^{2(\mbox{\^v}-1)}$ \hspace{0.2cm}
(\^v) \hspace{0.6cm}
${x}^{\mbox{\^v}}$ \hspace{0.4cm}
(\^v) \hspace{0.4cm} 
${x}^{2(\mbox{\^v}-1)}$ \hspace{0.2cm}
(\^v)\\  
$\cdots$
$\, \,$
(\^v) \hspace{0.4cm}
${x}^{2(\mbox{\^v}-1)}$ \hspace{0.1cm}
(\^v) \hspace{0.5cm}
${x}^{\mbox{\^v}}$ \hspace{0.6cm}
(\^v) \hspace{0.4cm}
${x}^{\mbox{\^v}-1}$\\
$\cdots$
$\, \,$ 
(\^v) \hspace{0.6cm}
${x}^{\mbox{\^v}}$ \hspace{0.6cm}
(\^v) \hspace{0.2cm}
${x}^{2(\mbox{\^v}-1)}$ \hspace{0.2cm}
(\^v) \hspace{0.6cm}
${x}^{\mbox{\^v}}$
\hspace{0.2cm} $\leftarrow$ blockage! \\
$\cdots$ 
${x}^{2(\mbox{\^v}-1)}$ \hspace{0.2cm}
(\^v) \hspace{0.6cm}
${x}^{\mbox{\^v}}$ \hspace{0.4cm}
(\^v) \hspace{0.4cm} 
${x}^{2(\mbox{\^v}-1)}$ \hspace{0.2cm}
(\^v) 
$\, \, \, \,$ blockage \\  
$\cdots$
$\, \,$
(\^v) \hspace{0.4cm}
${x}^{2(\mbox{\^v}-1)}$ \hspace{0.1cm}
(\^v) \hspace{0.5cm}
${x}^{\mbox{\^v}}$ \hspace{0.6cm}
(\^v) \hspace{0.4cm}
${x}^{\mbox{\^v}-1}$ 
$\, \, \, \, \,$ has no \\
$\cdots$
$\, \,$ 
(\^v) \hspace{0.6cm}
${x}^{\mbox{\^v}}$ \hspace{0.6cm}
(\^v) \hspace{0.2cm}
${x}^{2(\mbox{\^v}-1)}$ \hspace{0.2cm}
(\^v) \hspace{0.6cm}
${x}^{\mbox{\^v}}$ 
$\, \, \, \, \, \, \,$ effect \\ 
\\
 
}

 \newpage
 
 \section*{FIGURE CAPTIONS}
 
 \noindent
 {\bf Fig 1a:}\\
 Phase diagram with density profiles for ${\mbox{v}}_{max}$ = 1 in
 dependence on the injection rate $\alpha$ and the extinction rate
 $\beta$ (according to [17]-[22])
 \\
 \\
 {\bf Fig 1b:}\\
 Phase diagram in dependence on the injection rate $\alpha$
 and the extinction rate $\beta$ (${\mbox{v}}_{max}$ = 2)
 \\
 \\
 {\bf Fig 1c:}\\
 Phase diagram in dependence on the injection rate $\alpha$
 and the extinction rate $\beta$ (${\mbox{v}}_{max}$ = 3)
 \\
 \\
 {\bf Fig 1d:}\\
 Phase diagram in dependence on the injection rate $\alpha$
 and the extinction rate $\beta$ (${\mbox{v}}_{max}$ = 5).
 Our investigations are focused on the cases $\beta$ = 1,
 $\alpha$ = 1, and $\beta$ = 1-$\alpha$ marked by dashed lines.
 \\
 \\
 {\bf Fig 2a:}\\
 Current for $\beta$ = 1 and the maximum velocities
 ${\mbox{v}}_{max}$ = 2, 3, ..., 10
 \\
 \\
 {\bf Fig 2b:}\\
 Current for $\alpha$ = 1 and the maximum velocities
 ${\mbox{v}}_{max}$ = 2, 3, ..., 10
 \\
 \\
 {\bf Fig 2c:}\\
 Global density for $\alpha$ = 1 and the maximum velocities
 ${\mbox{v}}_{max}$ = 2, 3, ..., 10
 \\
 \\
 {\bf Fig 3a:}\\
 Current in dependence on the injection rate $\alpha$
 and the extinction rate $\beta$ (${\mbox{v}}_{max}$ = 5)
 \\
 \\
 {\bf Fig 3b:}\\
 Global density in dependence on the injection rate $\alpha$
 and the extinction rate $\beta$ (${\mbox{v}}_{max}$ = 5)
 \\
 \\
 {\bf Fig 4a:}\\
 Density profiles at the beginning of the system ($\beta$ = 1)
 \\
 \\
 {\bf Fig 4b:}\\
 Density profiles at the end of the system ($\beta$ = 1)
 \\
 \\
 {\bf Fig 5a:}\\
 Density profiles for $\alpha$ = 1
 \\
 \\
 {\bf Fig 5b:}\\
 Density profiles for $\alpha$ = 1 around the critical extinction rate
 ${\beta}_{c}$
 \\
 \\
 {\bf Fig 5c:}\\
 Detail from Fig 5b at the beginning of the system ($\alpha$ = 1)
 \\
 \\
 {\bf Fig 5d:}\\
 Detail from Fig 5b at the end of the system ($\alpha$ = 1)
 \\
 \\
 {\bf Fig 6a:}\\
 Density profiles from Fig 5b taking only the sites i = 6+5n into
 account ($\alpha$ = 1)
 \\
 \\
 {\bf Fig 6b:}\\
 Logarithmic plot of the density profiles for $\beta$ $>$ ${\beta}_{c}$
 taking only the sites i = 6+5n into account ($\alpha$ = 1)
 \\
 \\
 {\bf Fig 6c:}\\
 Logarithmic plot of ${\rho}_{max}$ - $\rho$(i=6+5n) for $\alpha$ = 1
 \\
 \\
 {\bf Fig 6d:}\\
 Gradient of the density profiles in Figs 6b,c depending on the extinction
 rate $\beta$ ($\alpha$ = 1)
 \\
 \\
 {\bf Fig 6e:}\\
 Maximum value of the density profiles $\rho$(i=6+5n) on i = 1021
 ($\alpha$ = 1)
 \\
 \\
 {\bf Fig 7a:}\\
 Comparison of the density profiles for $\beta$ = 1-$\alpha$ and $\beta$ = 1
 \\
 \\
 {\bf Fig 7b:}\\
 Comparison of the density profiles for $\beta$ = 1-$\alpha$ and $\alpha$ = 1
 \\
 \\
 {\bf Fig 8a:}\\
 Correlation functions for systems with periodic boundary conditions
 \\
 \\
 {\bf Fig 8b:}\\
 Correlation functions in the middle of the system for $\beta$ = 1-$\alpha$
 \\
 \\
 {\bf Fig 8c:}\\
 Correlation functions at the beginning of the system for $\beta$ = 1-$\alpha$
 \\
 \\
 {\bf Fig 8d:}\\
 Correlation functions at the end of the system for $\beta$ = 1-$\alpha$

\begin{thebibliography}{4}
 \bibitem {Zia:Sha:Schm:Ast}
 R. K. P. Zia, B. Shaw, B. Schmittmann, and R. J. Astalos,
 {\it Phys. Rep.} {\bf 301}, 45 (1998).
 %
 \bibitem {Schm:Zia}
 B. Schmittmann and R. P. K. Zia,
 {\it cond-math/980392}.
 %
 \bibitem {Kr}
 J. Krug, {\it Phys. Rev. Lett.} {\bf 67}, 1882 (1991).
 %
 \bibitem {Kr:Fer}
 J. Krug and P. A. Ferrari,
 {\it J. Phys. A} {\bf 29}, L465 (1996).
 %
 \bibitem {De:Do:Mu}
 B. Derrida, E. Domany, and D. Mukamel,
 {\it J. Stat. Phys.} {\bf 69}, 667 (1992).
 %
 \bibitem {De:Ev}
 B. Derrida and M. R. Evans,
 {\it J. Phys. I} {\bf 3}, 311 (1993).
 %
 \bibitem {De:Ev:Ha:Pa}
 B. Derrida, M. R. Evans, V. Hakim, and V. Pasquier,
 {\it J. Phys. A} {\bf 26}, 1493 (1993).
 %
 \bibitem {Schue:Do}
 G. Sch\"utz and E. Domany,
 {\it J. Stat. Phys.} {\bf 72}, 277 (1993).
 %
 \bibitem{Ra:Schr}
 N. Rajewsky and M. Schreckenberg,
 {\it Physica A} {\bf 245}, 139 (1997).
 %
 \bibitem{Schue}
 G. Sch\"utz,
 {\it Phys. Rev. E} {\bf 47}, 4265 (1993).
 %
 \bibitem{Hin}
 H. Hinrichsen,
 {\it J. Phys. A} {\bf 29}, 3659 (1996).
 %
 \bibitem{Hon:Pesch}
 A. Honecker and I. Peschel,
 {\it J. Stat. Phys.} {\bf 88}, 319 (1997).
 %
 \bibitem{Ra:Scha:Schr}
 N. Rajewsky, A. Schadschneider, and M. Schreckenberg,
 {\it J. Phys. A} {\bf 29}, L305 (1996).
 %
 \bibitem{Ti:Er}
 L. G. Tilstra and M. H. Ernst,
 {\it J. Phys. A} {\bf 31}, 5033 (1998).
 %
 \bibitem{Ev:Ra:Spe}
 M. R. Evans, N. Rajewsky, and E. R. Speer,
 {\it J. Stat. Phys.} {\bf 95}, 45 (1999).
 %
 \bibitem{Fu:Bo}
 H. Fuk\'s and N. Boccara,
 {\it Int. J. Mod. Phys. C} {\bf 9}, 1 (1998).
 %
 \bibitem{Fu}
 H. Fuk\'s, {\it Phys. Rev. E} {\bf 60}, 197 (1999).
 %
 \bibitem{Gi:Ni}
 J. de Gier and B. Nienhuis,
 {\it Phys. Rev. E} {\bf 59}, 4899 (1999).
 %
 \bibitem{Be:Cha:Ez}
 A. Benyoussef, H. Chakib, and H. Ez-Zahraouy,
 {\it Eur. Phys. J. B} {\bf 8}, 275 (1999).
 %
 \bibitem {Wo:Schr:Ba}
 Traffic and Granular Flow.
 Eds: D.E. Wolf, M. Schreckenberg, and A. Bachem,
 World Scientific (Singapore 1996).
 %
 \bibitem {Wo:Schr}
 Traffic and Granular Flow '97.
 Eds: M. Schreckenberg and D.E. Wolf, Springer (Singapore 1998).
 %
 \bibitem {Chowd:Sa:Scha}
 D. Chowdhury. L. Santen, and A. Schadschneider,
 Phys. Rep. {\bf 329,} 199 (2000).
 %
 \bibitem{Ra:Sa:Scha:Schr}
 N. Rajewsky, L. Santen, A. Schadschneider, and M. Schreckenberg
 {\it J. Stat. Phys.} {\bf 92}, 151 (1998).
 %
 \bibitem {Na:Schr}
 K. Nagel and M. Schreckenberg,
 {\it J. Phys. I France} {\bf 2}, 2221 (1992).
 %
 \bibitem{Schr:Scha:Na:Ito}
 M. Schreckenberg, A. Schadschneider, K. Nagel, and N. Ito,
 {\it Phys. Rev. E} {\bf 51}, 2939 (1995).
 %
 \bibitem {Sas:Kert}
 M. Sasv\'ari and J. Kert\'esz, {\it Phys. Rev. E} {\bf 56}, 4104 (1997).
 %
 \bibitem {Eis:Sa:Scha:Schr}
 B. Eisenbl\"atter, L. Santen, A. Schadschneider, and M. Schreckenberg,
 {\it Phys. Rev. E} {\bf 57}, 1309 (1998).
 %
 \bibitem {Chey:Kert:Schr}
 S. Cheybani, J. Kert\'esz, and M. Schreckenberg,
 {\it J. Phys. A} {\bf 31}, 9787 (1998).
 %
 \bibitem {Sa:Scha}
 L. Santen and A. Schadschneider,
 {\it cond-math/9711261}.
 %
 \bibitem{Ro:Lue:Us}
 L. Roters, S. L\"ubeck, and K. D. Usadel,
 {\it Phys. Rev. E} {\bf 59}, 2672 (1999).
 %
 \bibitem{Scha1:Schr1}
 A.Schadschneider and M. Schreckenberg,
 {\it J. Phys. A} {\bf 26}, L679 (1993).
 %
 \bibitem{Scha2:Schr2}
 A.Schadschneider and M. Schreckenberg,
 {\it J. Phys. A} {\bf 30}, L69 (1997).
 %
 \bibitem{Scha3:Schr3}
 A.Schadschneider and M. Schreckenberg,
 {\it J. Phys. A} {\bf 31}, L225 (1998).
 %
 \bibitem{Scha}
 A.Schadschneider,
 {\it Eur. Phys. J. B} {\bf 10}, 573 (1999).
 %
 \end{thebibliography}
 
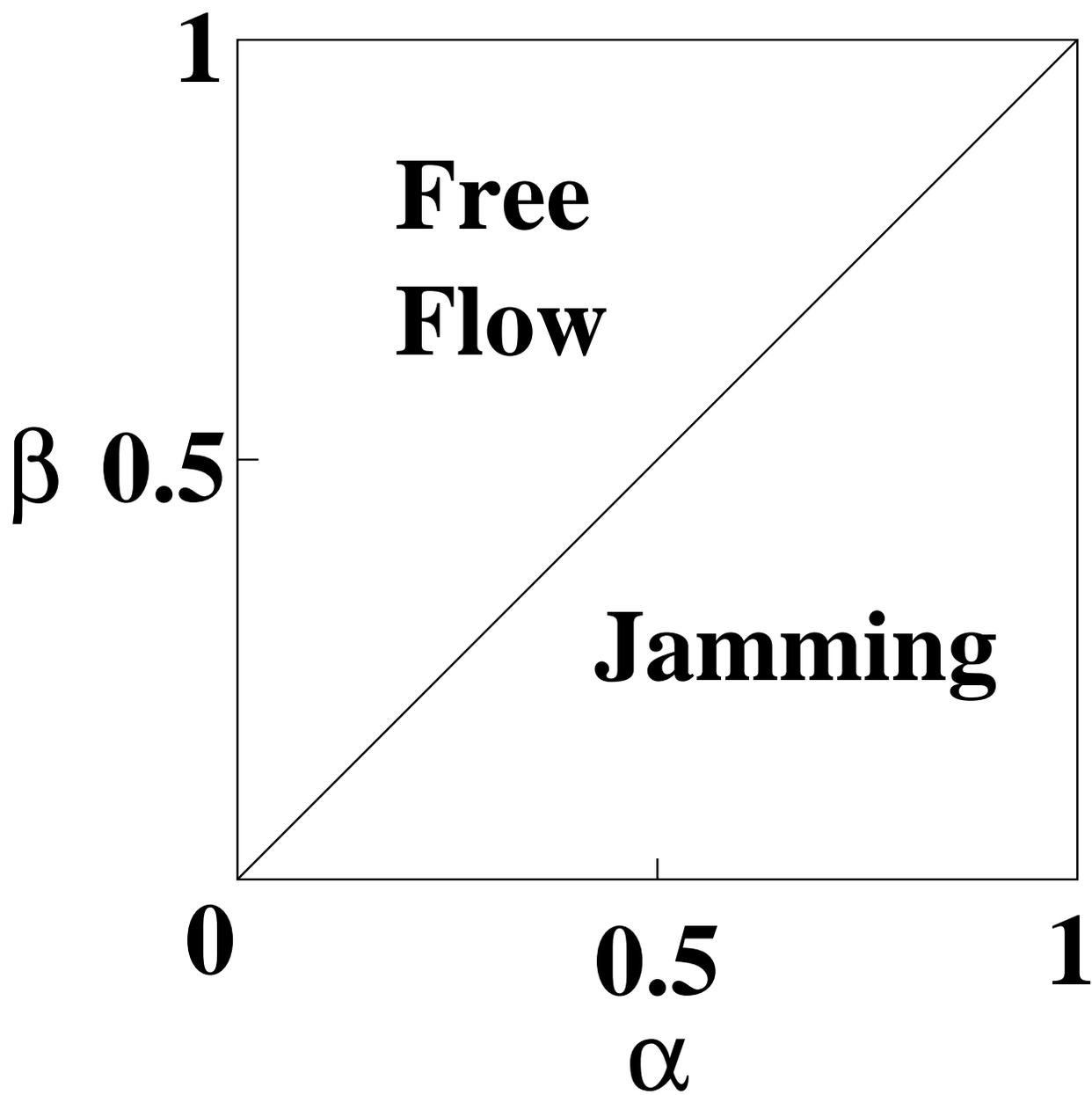
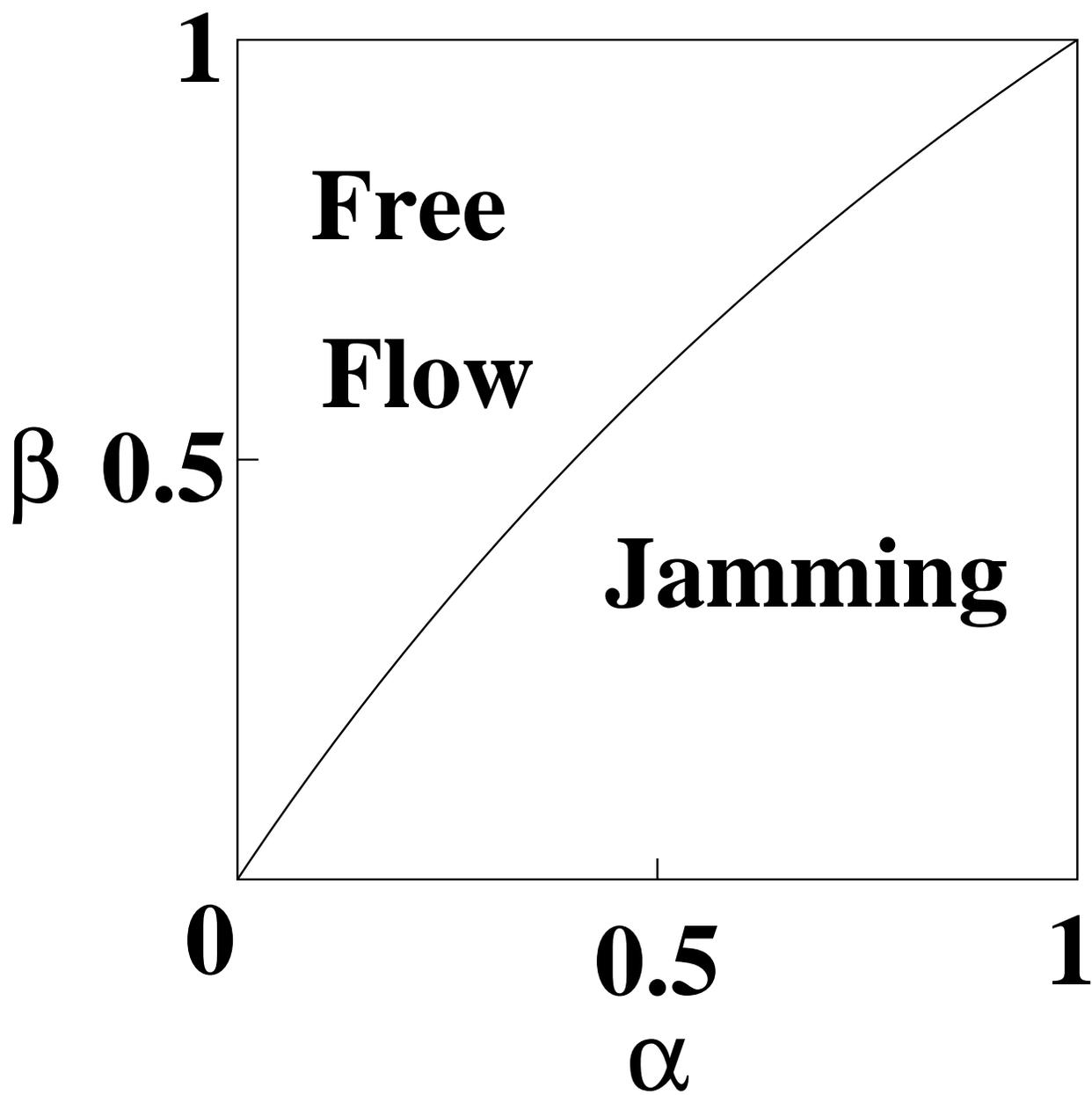
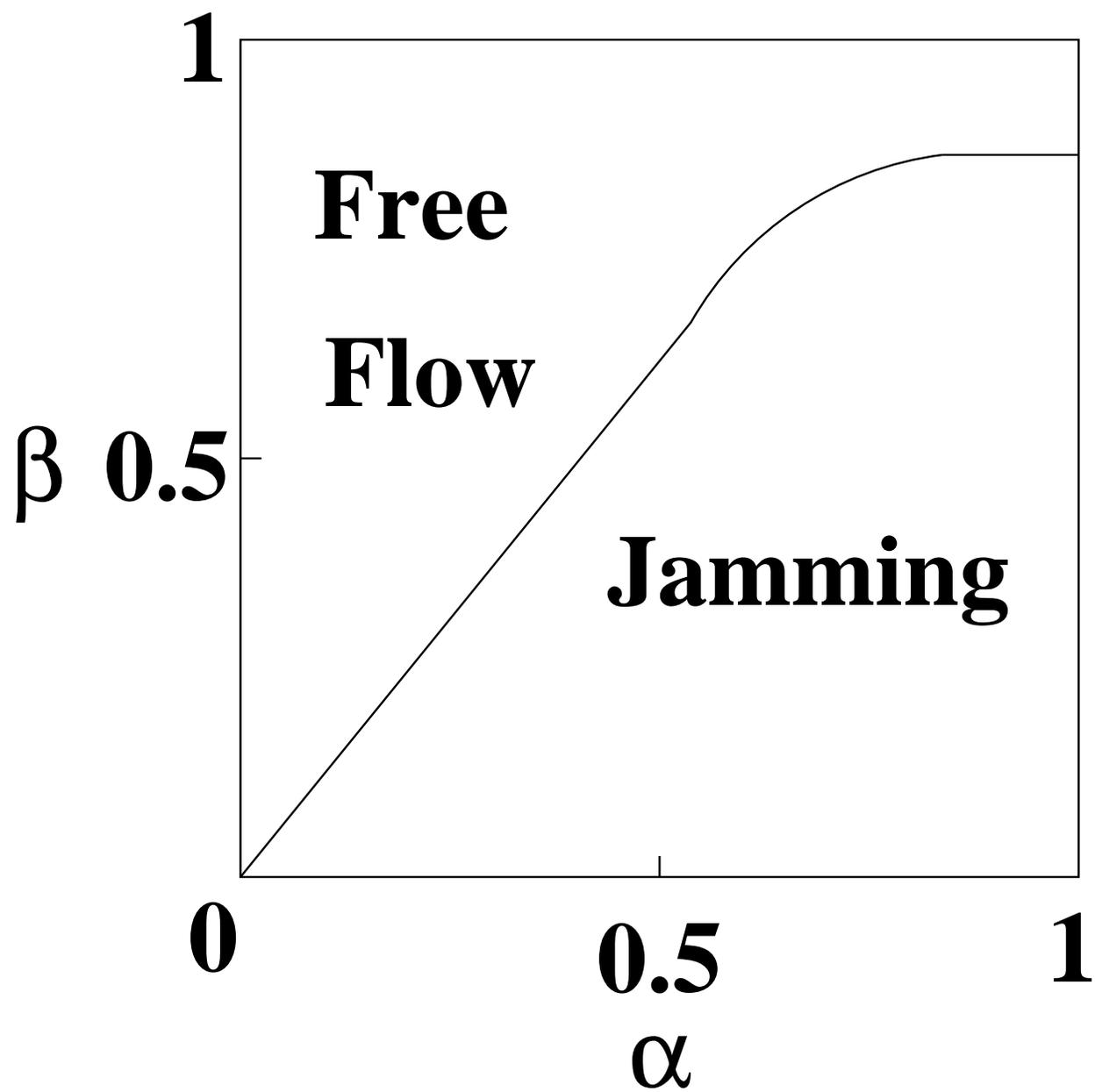
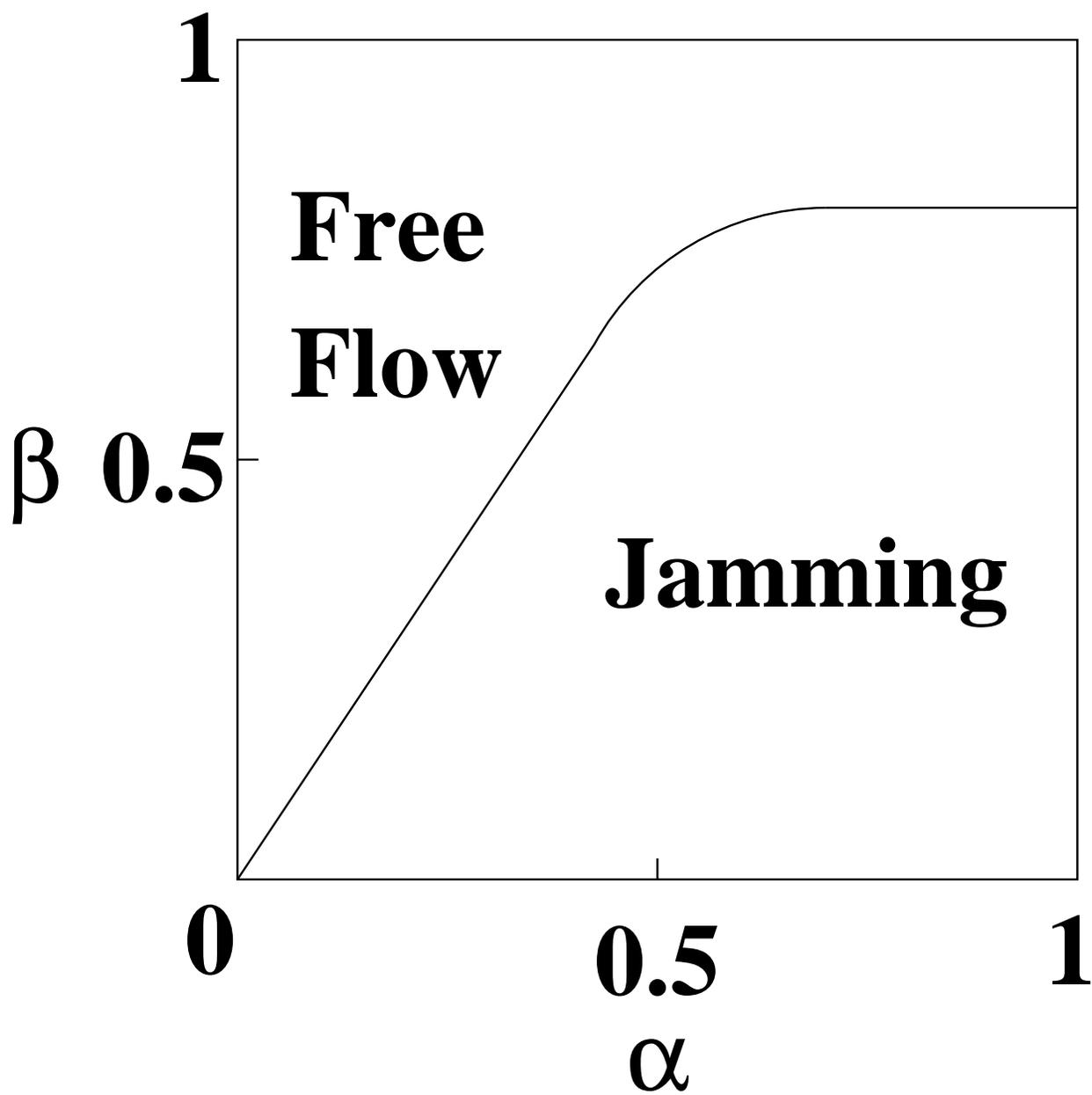
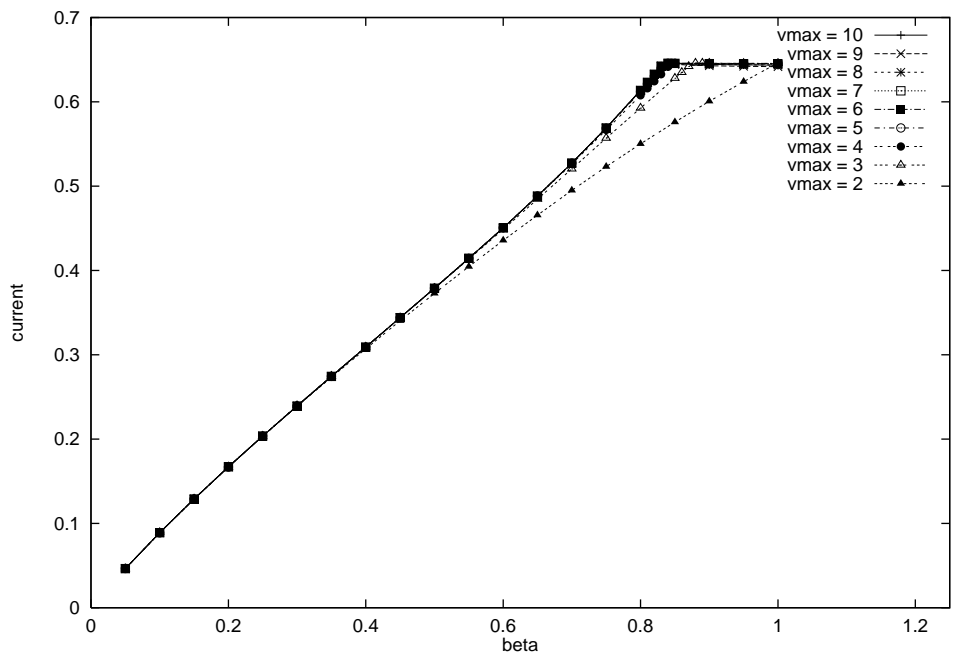
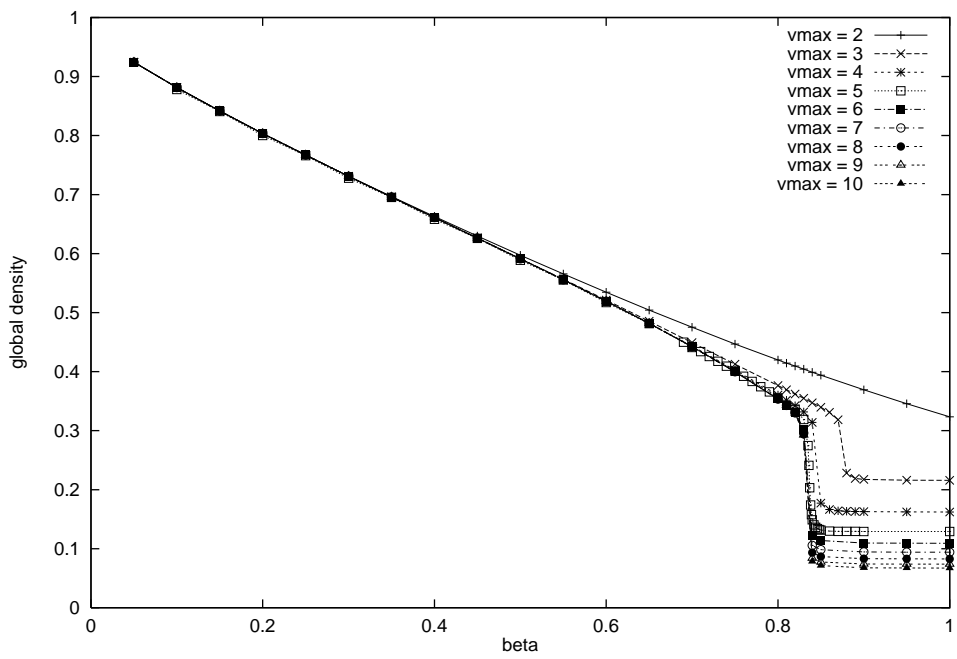
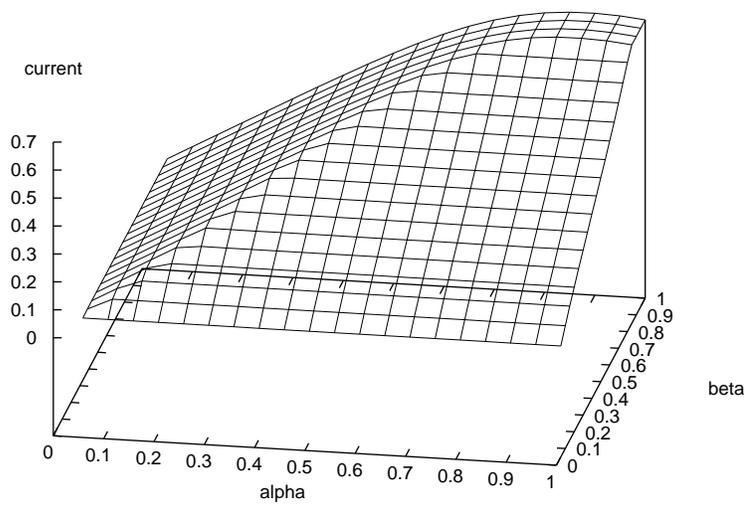
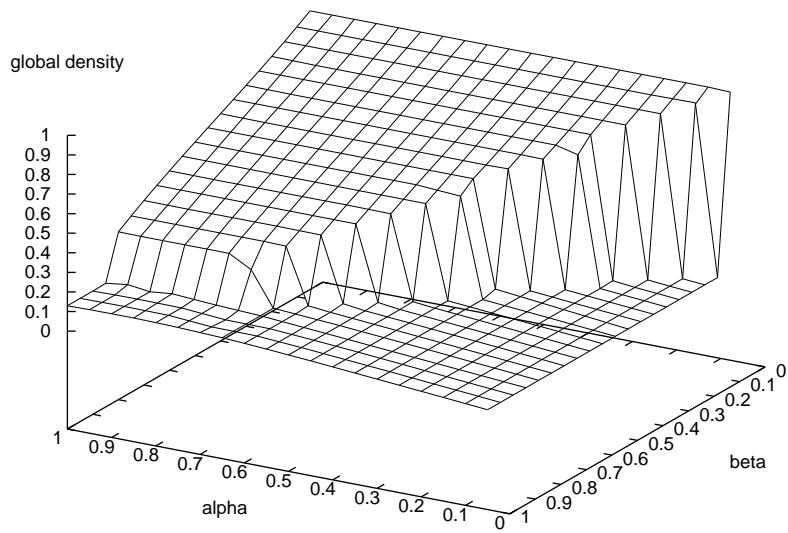
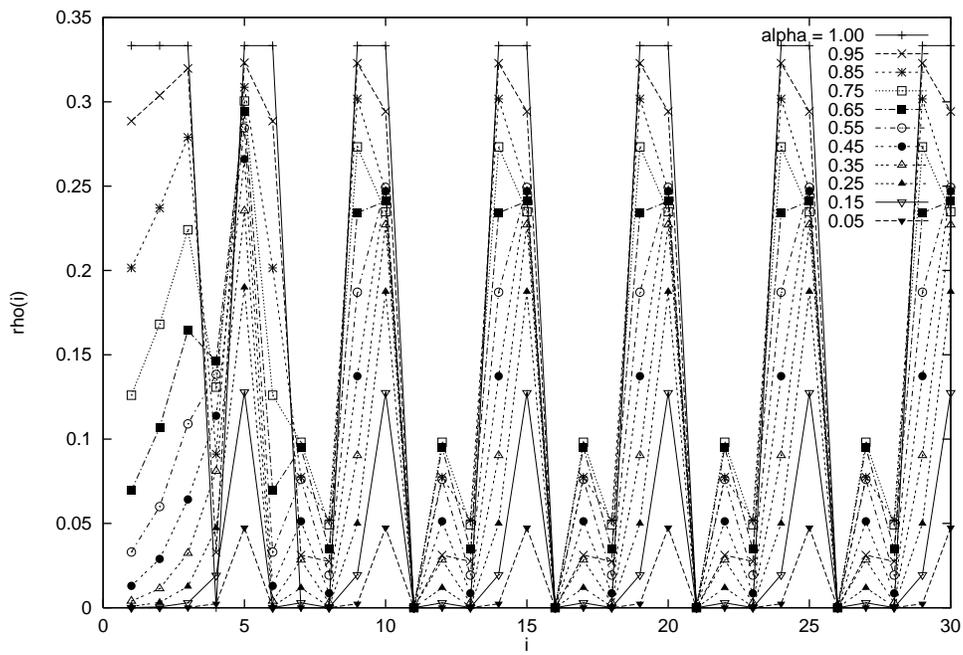
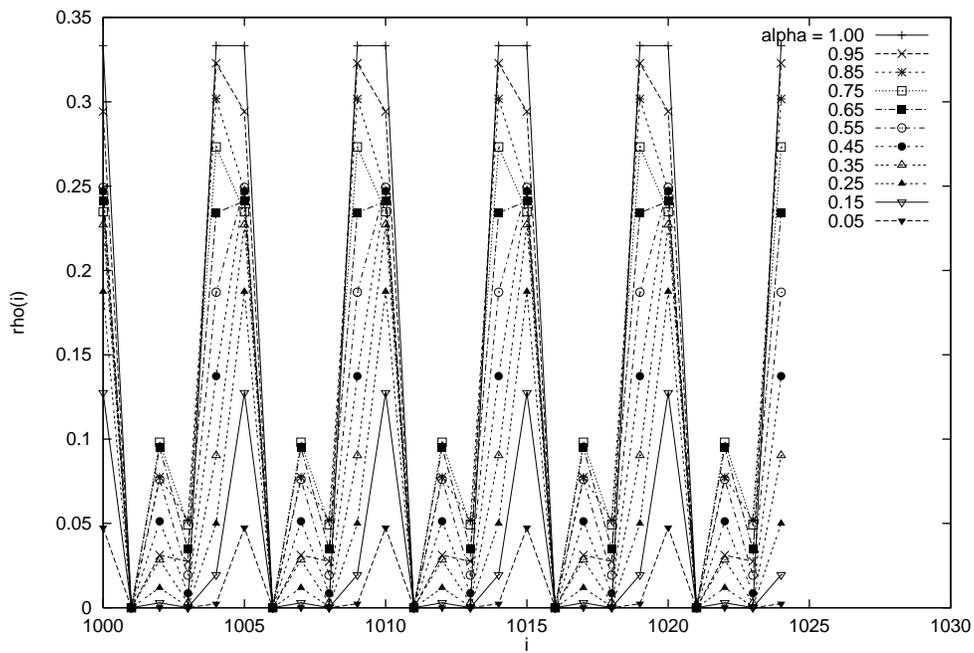
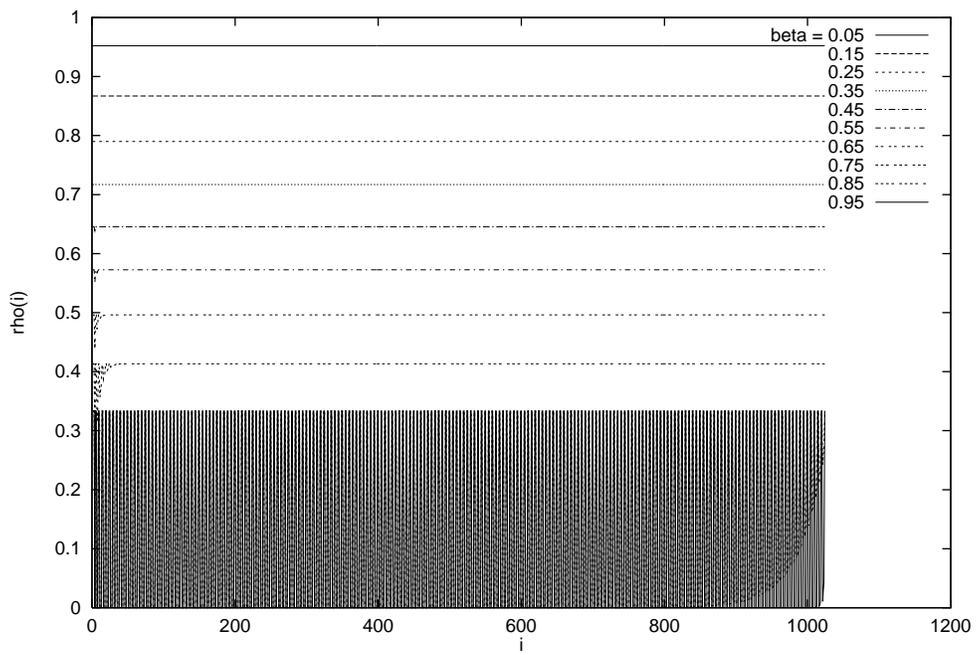
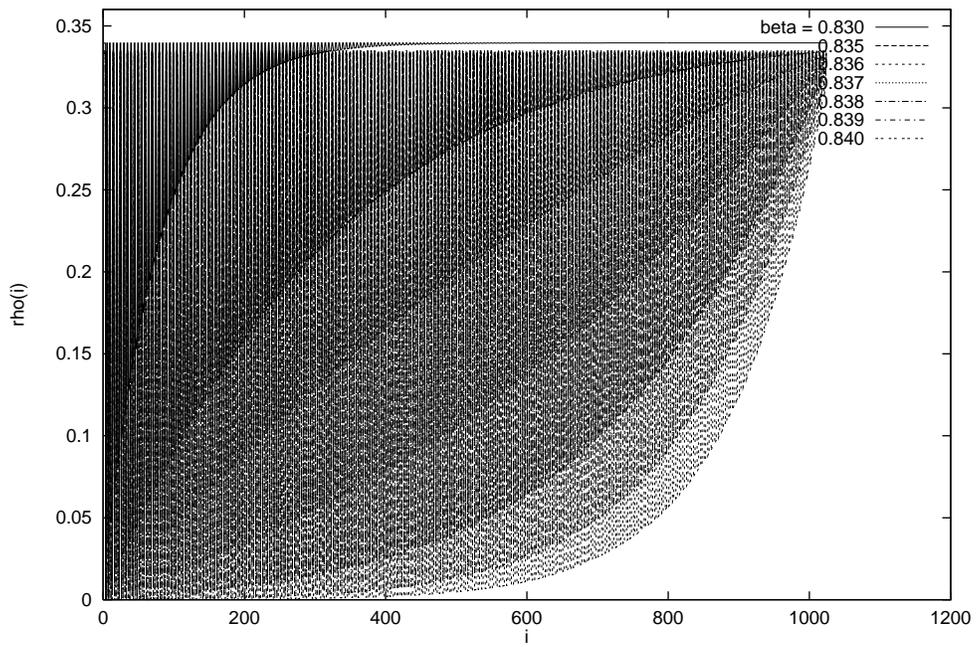
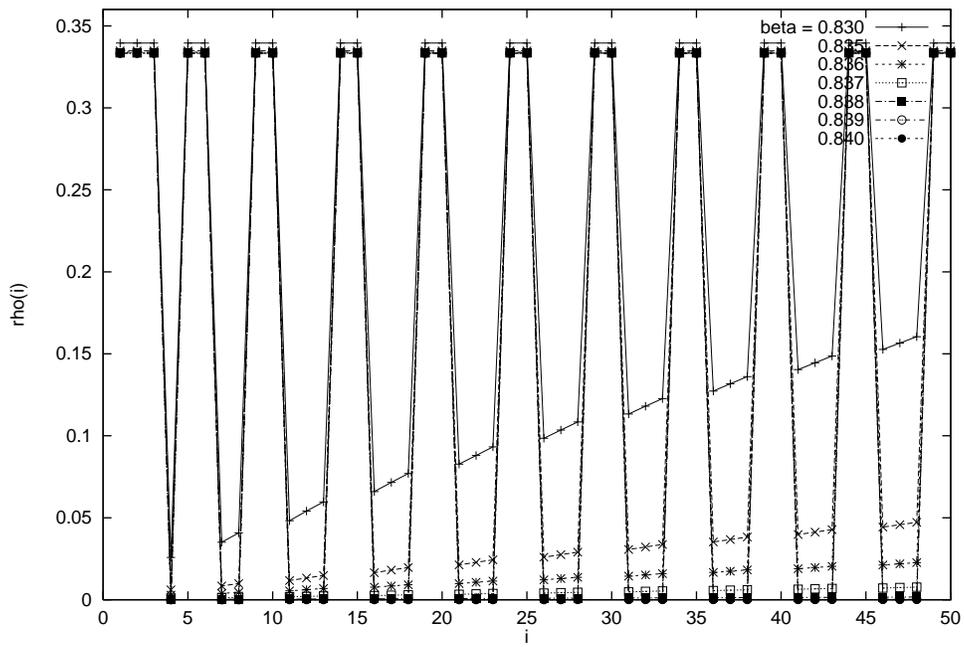
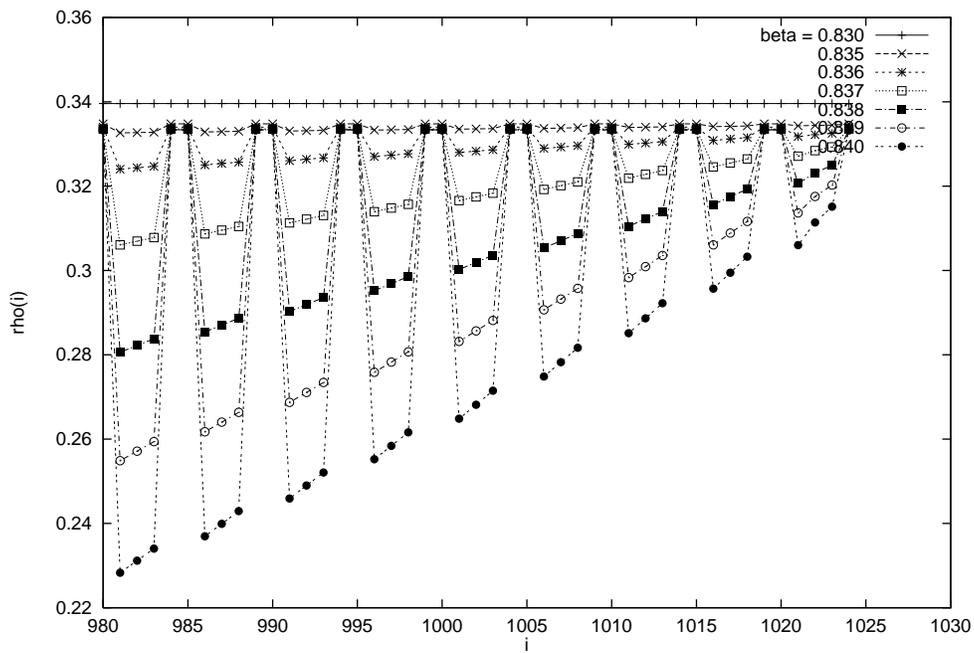
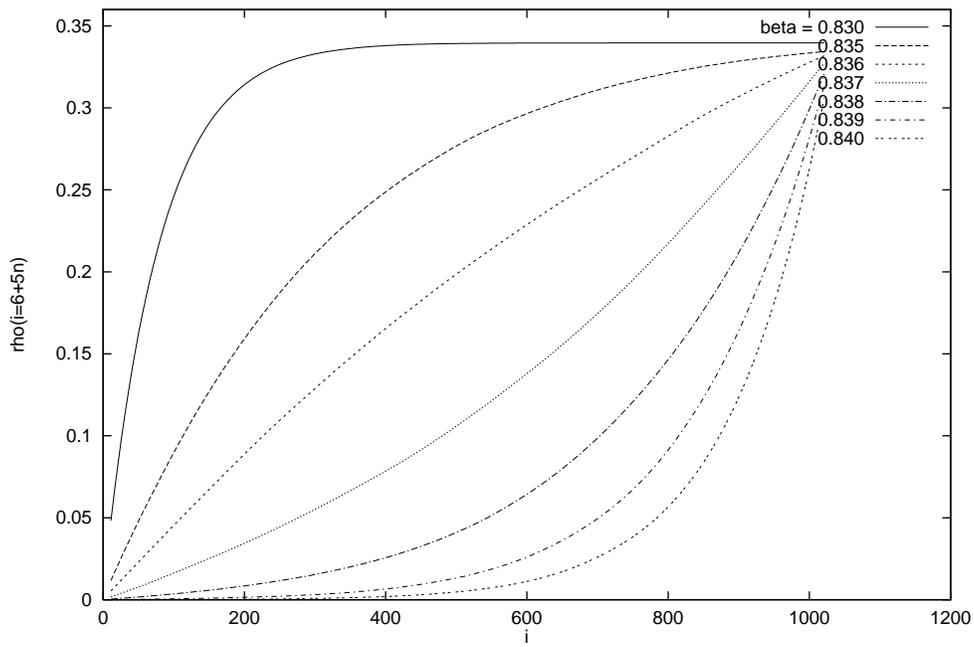
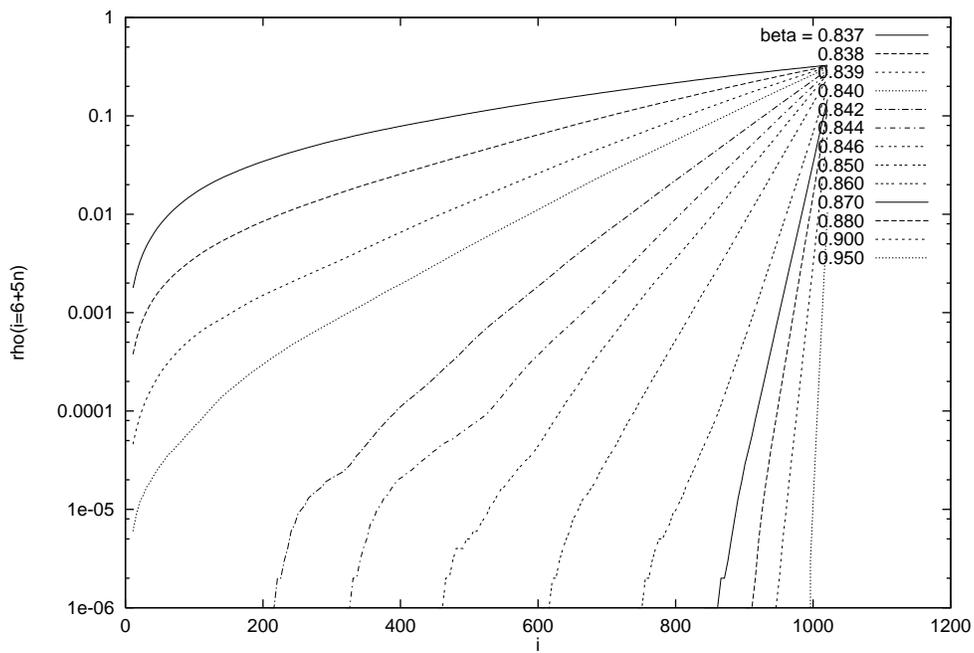
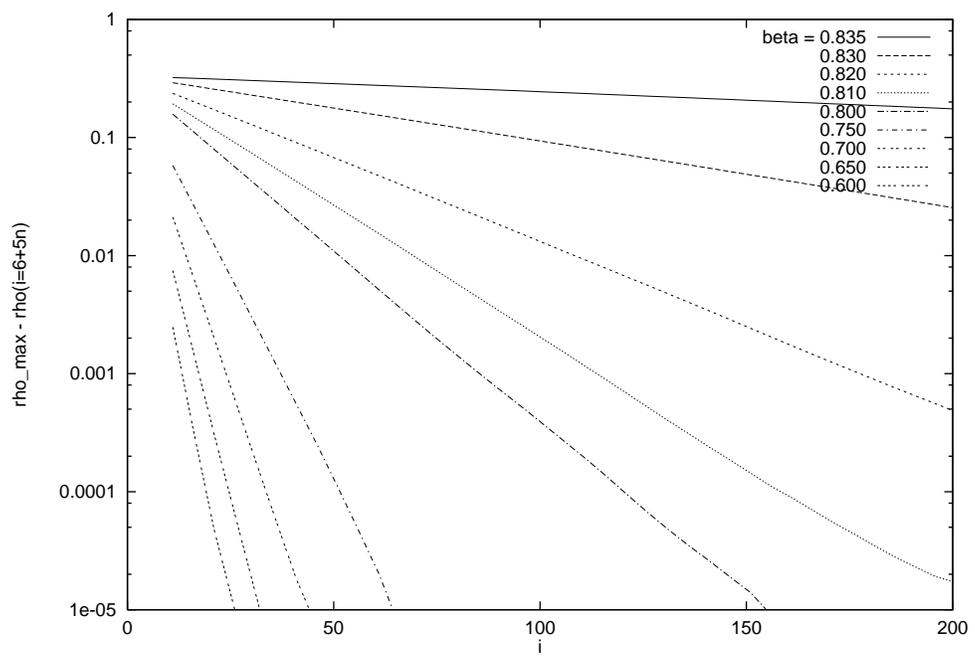
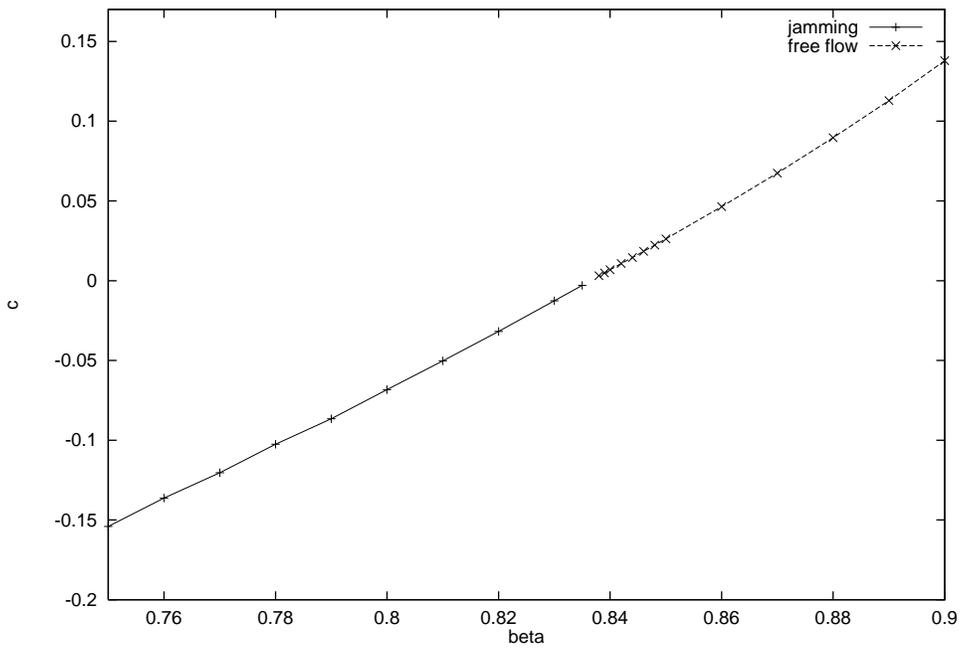
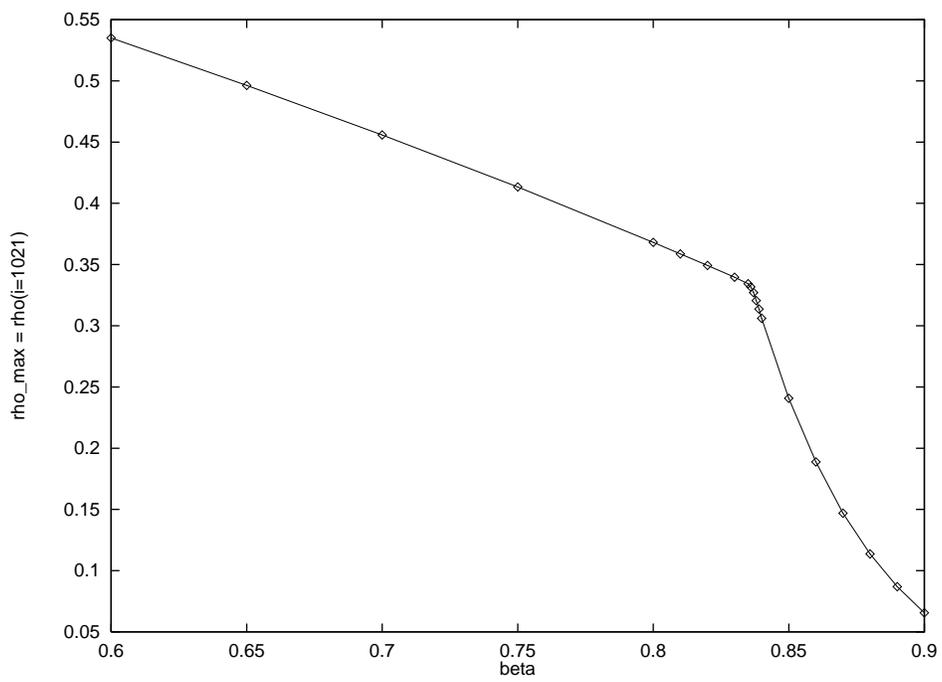
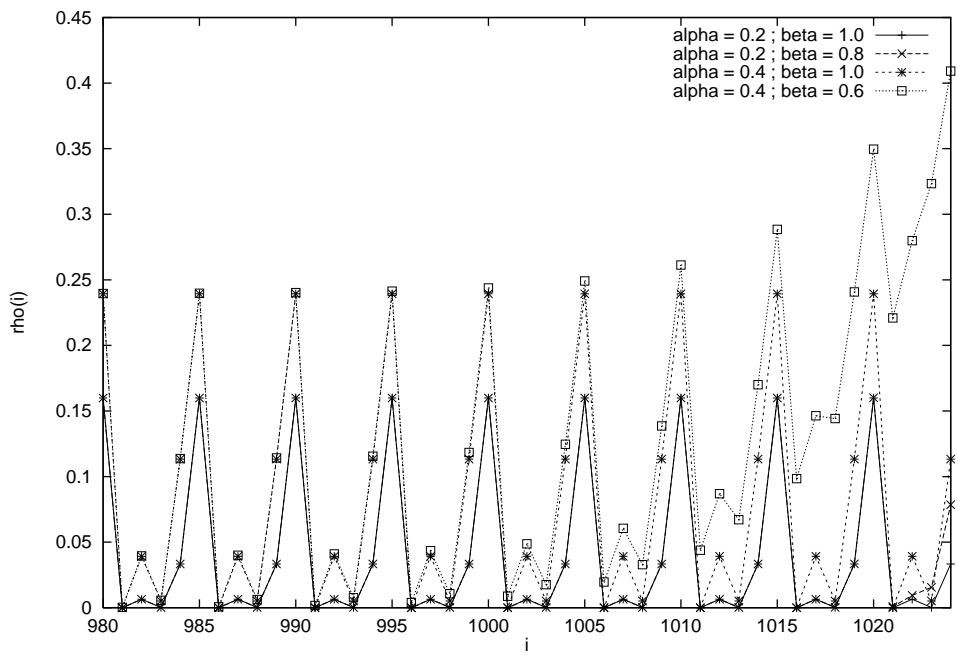
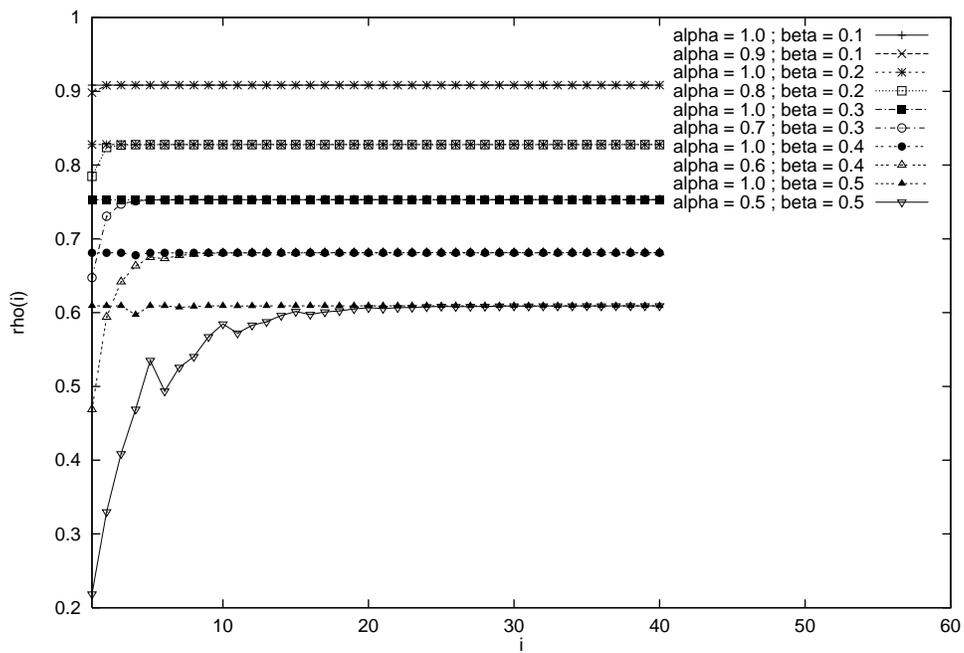
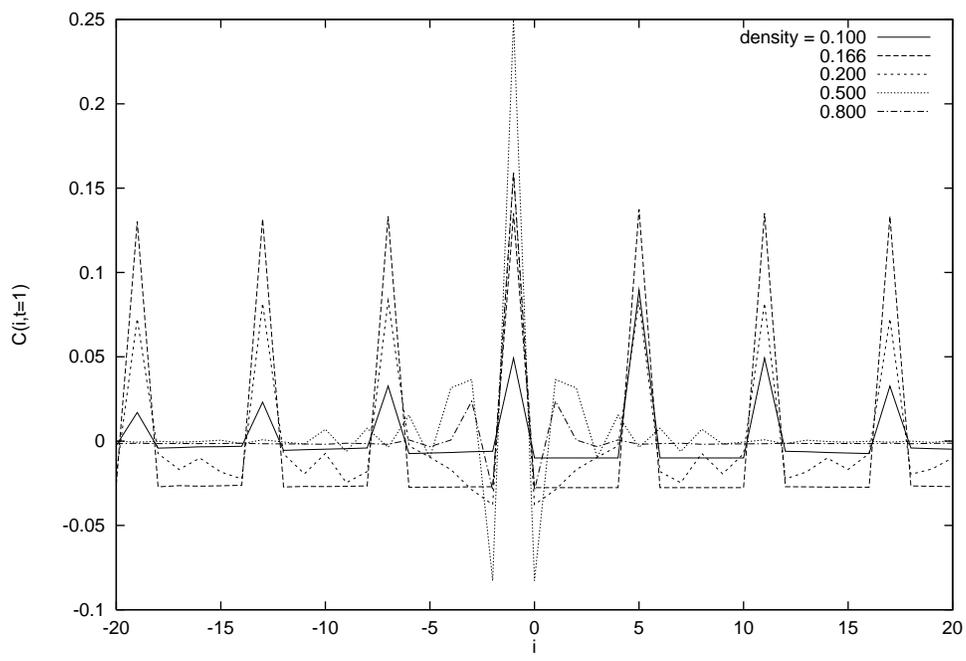
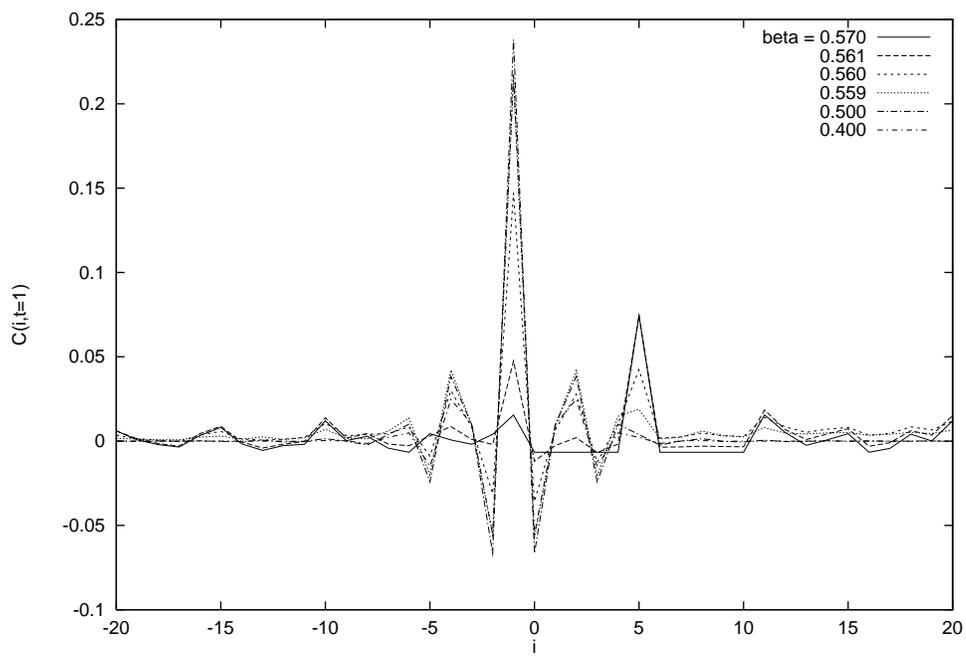
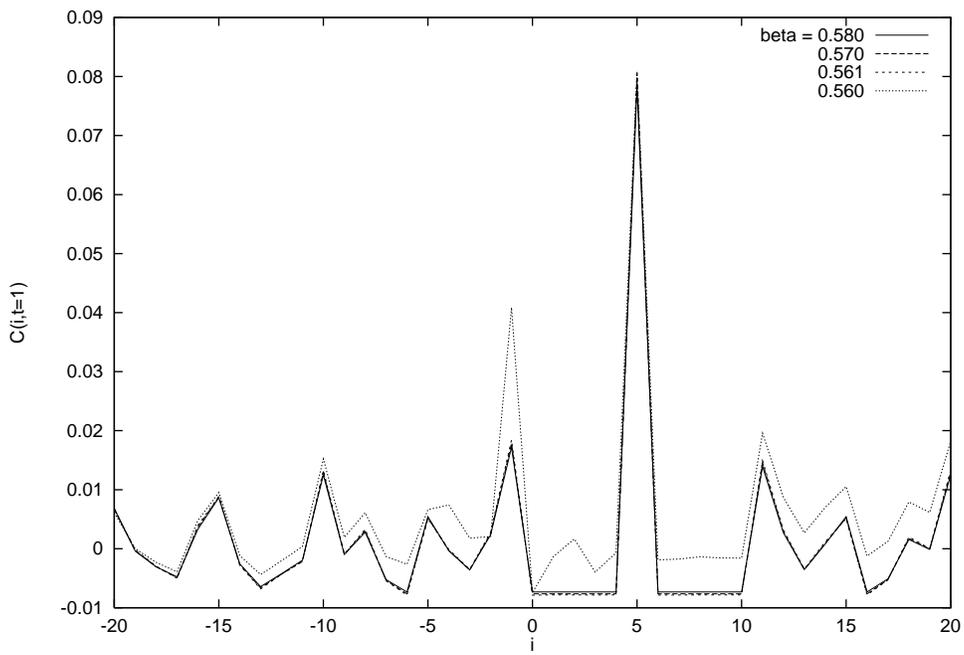
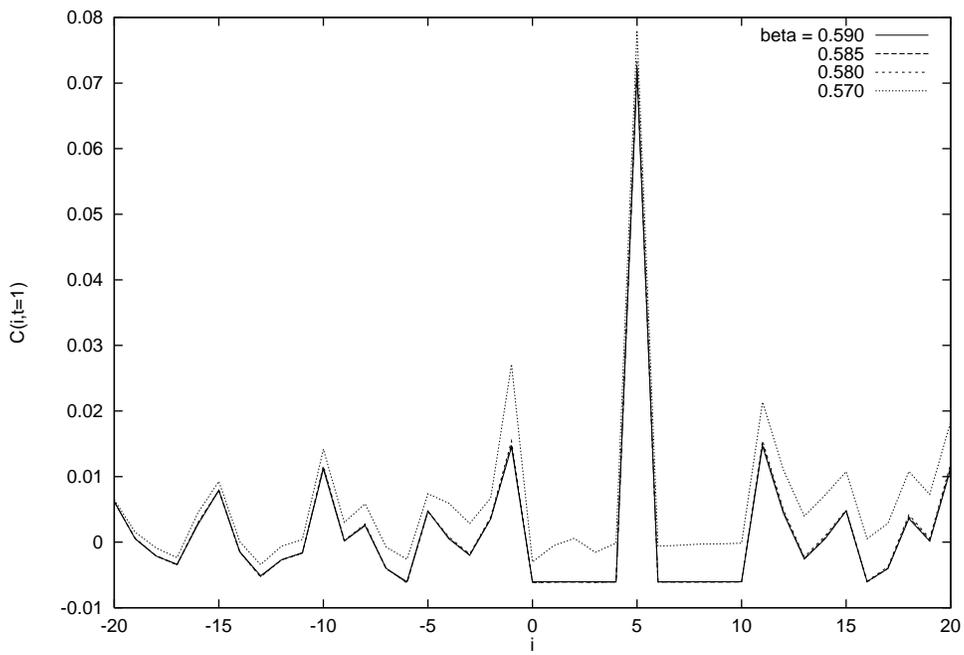
\end{document}